\documentclass[a4paper,11pt]{article}
\pdfoutput=1

\usepackage{amsmath,amssymb,bbm,mathtools}
\usepackage{ascmac}
\usepackage{cancel}
\usepackage{xcolor} 
\usepackage{graphicx} 
\usepackage{arydshln}
\usepackage{multirow}
\usepackage{marvosym,wasysym}
\newdimen\tbaselineshift 
\usepackage[bookmarks=true,bookmarksnumbered=true,bookmarkstype=toc]{hyperref} 
\hypersetup{
pdfauthor={Tetsuji KIMURA},
colorlinks={true},
linkcolor={blue},
urlcolor={blue},
filecolor={blue},
citecolor={blue}
}
\definecolor{refkey}{rgb}{0.40, 0.55, 0.55}
\definecolor{labelkey}{rgb}{0.40, 0.55, 0.55}
\usepackage{ulem}

 \font\f=cmr10
 \pdffontexpand\f 30 20 10 autoexpand
\makeatletter

 \@addtoreset{equation}{section}
\makeatother

\definecolor{darkblue}{HTML}{00008B}
\definecolor{brown}{HTML}{A52A2A}
\definecolor{darkcyan}{HTML}{008B8B}
\definecolor{darkgreen}{HTML}{006400}
\definecolor{darkolivegreen}{HTML}{556B2F}
\definecolor{deeppink}{HTML}{FF1493}
\definecolor{deepskyblue}{HTML}{00BFFF}
\definecolor{dodgerblue}{HTML}{1E90FF}
\definecolor{gray}{HTML}{808080}
\definecolor{hotpink}{HTML}{FF69B4}
\definecolor{lightblue}{HTML}{ADD8E6}
\definecolor{lightcyan}{HTML}{E0FFFF}
\definecolor{lightgray}{HTML}{D3D3D3}
\definecolor{lightgrey}{HTML}{D3D3D3}
\definecolor{lightpink}{HTML}{FFB6C1}
\definecolor{navy}{HTML}{000080}
\definecolor{orange}{HTML}{FFA500}
\definecolor{orchid}{HTML}{DA70D6}
\definecolor{pink}{HTML}{FFC0CB}
\definecolor{purple}{HTML}{800080}
\definecolor{royalblue}{HTML}{4169E1}
\definecolor{skyblue}{HTML}{87CEEB}
\definecolor{steelblue}{HTML}{4682B4}

\makeatletter
\def\tbcaption{\def\@captype{table}\caption}
\def\figcaption{\def\@captype{figure}\caption}
\makeatother

\newcounter{Enumerate}

\DeclareFontFamily{U}{rsf}{}
\DeclareFontShape{U}{rsf}{m}{n}{
  <5> <6> rsfs5 <7> <8> <9> rsfs7 <10-> rsfs10}{}
\DeclareMathAlphabet\Scr{U}{rsf}{m}{n}

\usepackage[mathscr]{eucal}

\newcommand{\LS}{\ \ \ \ \ \ \ \ \ \ }
\newcommand{\ls}{\ \ \ \ \ }
\newcommand{\wt}{\widetilde}

\newcommand{\bsubeq}{\begin{subequations}}
\newcommand{\esubeq}{\end{subequations}}

\newcommand{\noi}{\noindent}

\newcommand{\A}{{\bf A}}
\newcommand{\B}{{\bf B}}
\newcommand{\C}{{\bf C}}

\newcommand{\E}{{\bf E}}
\newcommand{\I}{{\rm i}}
\newcommand{\N}{\mathcal{N}}
\newcommand{\T}{{\rm T}}
\newcommand{\X}{{\bf X}}

\renewcommand{\d}{{\rm d}}
\newcommand{\e}{{\rm e}}
\renewcommand{\l}{\ell}

\newcommand{\slb}{\scalebox}

\newcommand{\txc}[1]{\textcircled{#1}}
\newcommand{\tcc}[1]{\textcolor{red}{$\bullet^{\sf #1}$}}

\begin{document}
\allowdisplaybreaks{

\thispagestyle{empty}


\begin{flushright}
TIT/HEP-651 
\end{flushright}

\vspace{35mm}

\noi
\slb{2.3}{Exotic Brane Junctions from F-theory}
%
%


\vspace{15mm}

\noi
{\renewcommand{\arraystretch}{1.6}
\begin{tabular}{cl}
\multicolumn{2}{l}{\slb{1.2}{Tetsuji {\sc Kimura}} \vphantom{$\Bigg|$}}
\\
& {\renewcommand{\arraystretch}{1.0}
\begin{tabular}{l}
{\sl Research and Education Center for Natural Sciences, Keio University}
\\
{\sl Hiyoshi 4-1-1, Yokohama, Kanagawa 223-8521, JAPAN} 
\end{tabular}
}
\\
& \ls and
\\
& {\renewcommand{\arraystretch}{1.0}
\begin{tabular}{l}
{\sl
Department of Physics,
Tokyo Institute of Technology} \vphantom{$\Big|$}
\\
{\sl Tokyo 152-8551, JAPAN}
\end{tabular}
}
\\
& \ \ \ \slb{0.9}{\tt tetsuji.kimura \_at\_ keio.jp}
\end{tabular}
}

\vspace{25mm}

\begin{abstract}
Applying string dualities to F-theory, we obtain various $[p,q]$-branes whose constituents are standard branes of codimension two and exotic branes.
We construct junctions of the exotic five-branes and their Hanany-Witten transitions associated with those in F-theory.
In this procedure, we understand the monodromy of the single $5^2_2$-brane. 
We also find the objects which are sensitive to the branch cut of the $5^2_2$-brane.
Considering the web of branes in the presence of multiple exotic five-branes analogous to the web of five-branes with multiple seven-branes, we obtain novel brane constructions for $SU(2)$ gauge theories with $n$ flavors and their superconformal limit with enhanced $E_{n+1}$ symmetry in five, four, and three dimensions.
Hence, adapting the techniques of the seven-branes to the exotic branes, 
we will be able to construct F-theories in diverse dimensions.
\end{abstract}


\newpage
\section{Introduction}
\label{sect:introduction}

What is the role of exotic branes in string theory?
The exotic branes appear when we apply string dualities such as T- and S-dualities to standard branes such as D-branes, NS5-brane, and fundamental string in lower dimensional spacetime \cite{Blau:1997du, Obers:1998fb, Eyras:1999at, LozanoTellechea:2000mc, deBoer:2010ud, Kikuchi:2012za}.
The features of the exotic branes are different from those of the standard branes.
First, the tension of the exotic brane is often stronger than those of the standard branes.
Second, the transverse dimensions of each exotic brane are two (or less).
Then the harmonic function which governs the background geometry of the exotic brane is given as a logarithmic function.
Strictly speaking, this implies that one cannot treat the exotic brane as a stand-alone object. 
Third, the configuration has a non-trivial monodromy associated with the string dualities.

As pointed in the above references, we have already studied string theories in terms of exotic branes.
In F-theory \cite{Greene:1989ya, Vafa:1996xn}, there exists a $[p,q]$ 7-brane.
This is an rotated object from a single D7-brane via the $SL(2,{\mathbb Z})_S$ S-duality.
In particular, when we define the $[1,0]$ 7-brane as a D7-brane, 
the $[0,1]$ 7-brane, often called the NS7-brane, is nothing but an exotic brane. 
This is referred to as a $7_3$-brane in the framework of the exotic branes. 
Furthermore, all of the exotic branes are descendants of the $7_3$-brane via the string dualities.

It has been investigated that there are non-trivial duality relations among various branes of codimension two.
Such branes are called {\it defect branes} \cite{Bergshoeff:2011se}.
They are governed by $SL(2,{\mathbb Z})_E$ duality transformation group. 
This duality group is called the {\it exotic duality} \cite{LozanoTellechea:2000mc, Bergshoeff:2011se, Sakatani:2014hba}.
This duality group is a subgroup of the U-duality group $E_{d(d)}({\mathbb Z})$ in $(11-d)$-dimensional spacetime with maximal supersymmetries.
In particular, 
the D7-brane and the exotic $7_3$-brane are subject to the $SL(2,{\mathbb Z})_S$ S-duality because of the equality $E_{1(1)}({\mathbb Z}) = SL(2,{\mathbb Z})_S$. 
This is also one representation of the exotic duality in string theory.
Performing the T-duality transformations along certain worldvolume directions of the seven-branes, 
one obtains various pairs of the $SL(2,{\mathbb Z})$ duality groups in diverse dimensions as in Table \ref{table:SL2s}:
\begin{table}[h]
\begin{center}
\slb{.9}{\renewcommand{\arraystretch}{1.3}
\begin{tabular}{c||rl|r@{\!\,\,\,}c@{\!\,\,\,}l} \hline
$D$ &
\multicolumn{2}{c|}{defect branes} & \multicolumn{3}{|c}{$SL(2,{\mathbb Z})$ duality group} 
\\ \hline\hline
10B & D7-brane, & $7_3$-brane 
& $SL(2,{\mathbb Z})_S$ &=& $E_{1(1)} ({\mathbb Z})$ 
\\
$p+3$ & defect D$p$-branes, & $p^{7-p}_3$-branes 
& $SL(2,{\mathbb Z})_E $ &$\subset$& $E_{8-p(8-p)}({\mathbb Z})$
\\
8 & defect NS5-brane, & $5^2_2$-brane 
& $SL(2,{\mathbb Z})_{\rho}$ &$\subset$& $SO(2,2;{\mathbb Z})_T$
\\
8 & \multicolumn{2}{c|}{defect (anti) KK5-branes} 
& $SL(2,{\mathbb Z})_{\tau}$ &$\subset$& $SO(2,2;{\mathbb Z})_T$ 
\\
4 & defect F-string, & $1^6_4$-brane 
& $SL(2,{\mathbb Z})_E$ &$\subset$& $E_{7(7)} ({\mathbb Z})$
\\
3 & defect pp-wave, & $0^{(1,6)}_4$-brane 
& $SL(2,{\mathbb Z})_E$ &$\subset$& $E_{8(8)} ({\mathbb Z})$
\\ \hline
\end{tabular}
}
\caption{Various pairs of the $SL(2,{\mathbb Z})$ duality groups in $D$-dimensional spacetime. $SO(2,2;{\mathbb Z})_T$ is the T-duality group on a compactified two-torus.}
\label{table:SL2s}
\end{center}
\end{table}

It is also useful to visualize the duality relations among various branes of codimension two in Figure \ref{fig:E-dualities}:
\begin{center}
\slb{.8}{\includegraphics[bb=0 0 509 260]{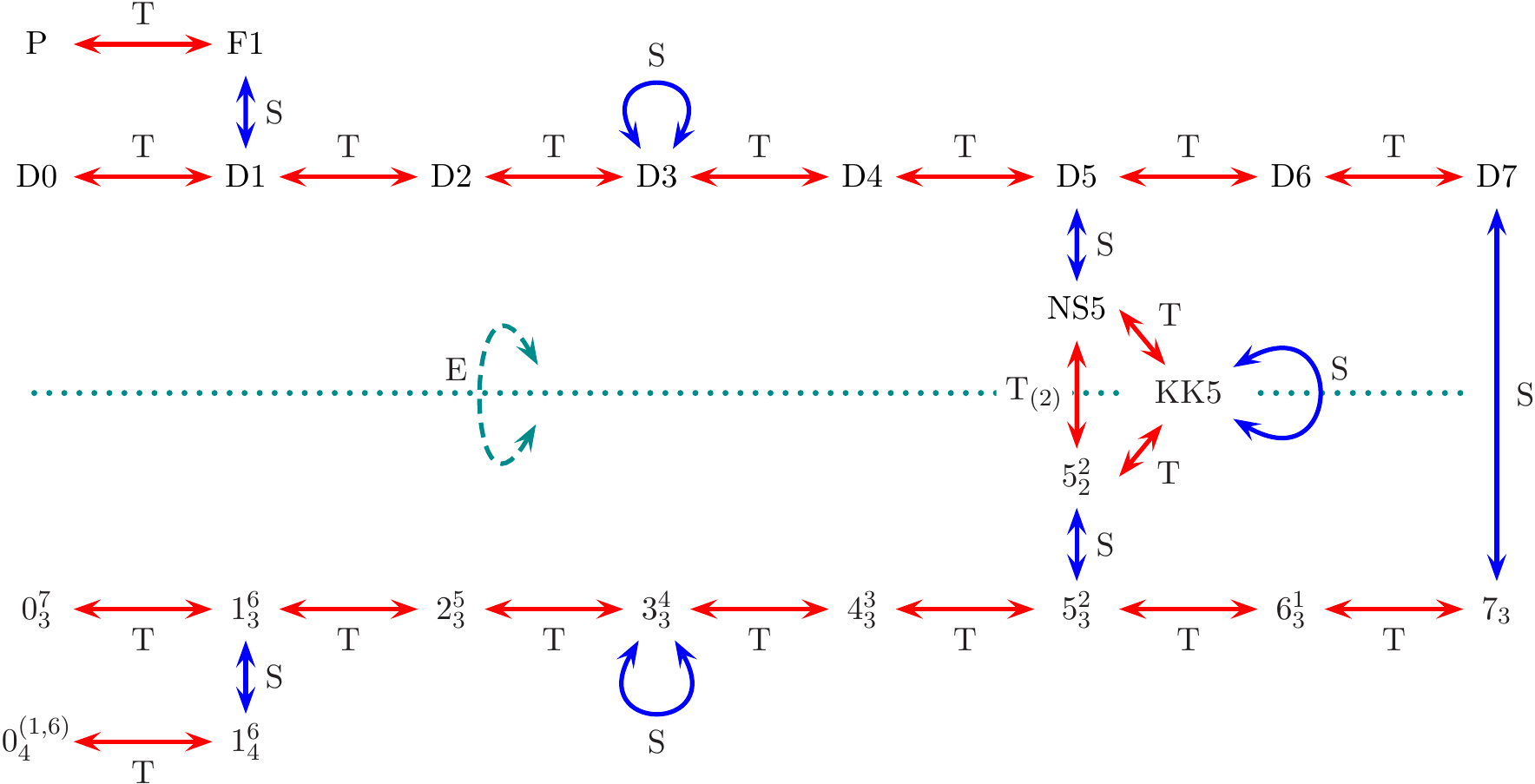}}
\figcaption{T-duality, S-duality and exotic duality (labelled as T, S and E, respectively) among the defect branes which contain the standard branes and the exotic branes. The label T$_{(2)}$ means the T-dualities along two directions.}
\label{fig:E-dualities}
\end{center}
As one can see immediately, each D$p$-brane in $(p+3)$-dimensional spacetime has an $SL(2,{\mathbb Z})_E$ pair such as a $p^{7-p}_3$-brane.
This exotic brane has a strong tension proportional to $g_s^{-3}$, where $g_s$ is the string coupling constant.

The author of this paper has studied mainly the exotic $5^2_2$-brane 
in the frameworks of the supergravity \cite{Kimura:2014wga, Kimura:2014bea, Kimura:2016xzd},
the worldvolume theory \cite{Kimura:2014upa, Kimura:2016anf},
and the gauged linear sigma model \cite{Kimura:2013fda, Kimura:2013zva, Kimura:2013khz, Kimura:2015yla, Kimura:2015cza, Kimura:2015qze}.
In these works, we mainly analyzed a {\it single} exotic $5^2_2$-brane.
In this paper, we investigate the branch cut of the single exotic brane \cite{Okada:2014wma}, the brane junctions by the cut \cite{Gaberdiel:1997ud, DeWolfe:1998zf, DeWolfe:1998yf, DeWolfe:1998eu, DeWolfe:1998pr}, and the Hanany-Witten transitions \cite{Hanany:1996ie}.
We further study the brane configurations \cite{Giveon:1998sr} which provide supersymmetric gauge theories on the branes \cite{Aharony:1997bh, DeWolfe:1999hj, Benini:2009gi, Bergman:2013ala, Bao:2013pwa, Hayashi:2013qwa, Kim:2014nqa, Kim:2015jba} in the presence of {\it multiple} exotic five-branes.
In particular, we extract the features of the exotic $5^2_2$-brane.

Since we discuss various configurations in this paper, we introduce the following nomenclature:
\begin{itemize}
\item $(p,q)$-string. 
This is an $SL(2,{\mathbb Z})_S$ S-dualized object from a fundamental string (or an F-string, for short) in ten-dimensional type IIB string theory, where $p$ and $q$ are relatively prime.
In other words, an F-string and a D-string are represented as the $(1,0)$-string and the $(0,1)$-string, respectively.
We sometimes refer to the $(p,q)$-string as the $(p,q)_1$-brane.

\item $(p,q)$ 5-brane. 
This is a five-brane on which the $(p,q)$-string is ending. 
This is also expressed as a $(p,q)_5$-brane in this work.
For instance, a D5-brane and an NS5-brane are given as the $(1,0)_5$-brane and the $(0,1)_5$-brane, respectively.

\item $(p,q)_b$-brane. 
This is an object extending in the $12\cdots b$-directions.
This originates from the $(p,q)_5$-brane via the string dualities.
We will remark its constituents case by case.

\item $[p,q]$ 7-brane. 
This is a seven-brane on which a $(p,q)$-string is ending in ten-dimensional type IIB string theory. 
We mainly refer to it as $[p,q]^S_7$-brane in this paper.
The superscript $S$ implies that this is subject to the $SL(2,{\mathbb Z})_S$ S-duality and the subscript 7 means the spatial dimensions.
Here the $[1,0]^S_7$-brane is nothing but a D7-brane, and the $[0,1]^S_7$-brane is an exotic $7_3$-brane. 

\item $[p,q]^E_{dn}$-brane ($0 \leq n \leq 7$). 
This is a defect brane of codimension two in $(n+3)$-dimensional spacetime. 
This originates from the $[p,q]^S_7$-brane via the T-duality transformations along the $(7-n)$ worldvolume directions.
The superscript $E$ denotes that this object is governed by the $SL(2,{\mathbb Z})_E$ exotic duality which is the subgroup of the U-duality group $E_{8-n(8-n)}({\mathbb Z})$ in $(n+3)$ dimensions.
The subscript $dn$ mentions that this object is a ``Dirichlet'' $n$-brane or its $SL(2,{\mathbb Z})_E$ exotic dualized one.
For example, the $[1,0]^E_{d5}$-brane is a defect D5-brane in eight dimensions, and $[0,1]^E_{d5}$-brane is an exotic $5^2_3$-brane, as in Figure \ref{fig:E-dualities}.
We remark that the $[p,q]^E_{d7}$-brane corresponds to the $[p,q]^S_7$-brane because of the equality $E_{1(1)}({\mathbb Z}) = SL(2,{\mathbb Z})_S$.

\item $[p,q]^T_{s5}$-brane. 
This is a defect five-brane in eight-dimensional theory. 
The superscript $T$ means that this object is given by the $SL(2,{\mathbb Z})_{\rho}$ duality group which is the part of the T-duality group $SO(2,2;{\mathbb Z})_T = SL(2,{\mathbb Z})_{\rho} \times SL(2,{\mathbb Z})_{\tau}$ coming from the compactified two-torus.
We note that $\rho$ and $\tau$ govern the background metric and the B-field on the torus and the complex structure of it, respectively.
The subscript $s5$ represents that this object is a ``solitonic'' five-brane whose tension is proportional to $g_s^{-2}$.
A $[1,0]^T_{s5}$-brane is a defect NS5-brane whose two of four transverse directions are compactified and smeared.
A $[0,1]^T_{s5}$-brane is an exotic $5^2_2$-brane.
Previously, the $[p,q]^T_{s5}$-brane is referred to as the ``defect $(p,q)$ five-brane'' in \cite{Kimura:2014wga}.
The $[p,q]^T_{s5}$-brane is obtained from the $[p,q]^S_7$-brane via the T-dualities along two transverse directions, and the S-duality.

Analogously, we also introduce a $[p,q]^T_{k5}$-brane.
This is a solitonic object whose constituents are ``Kaluza-Klein'' (KK) monopoles of codimension two, referred to as defect KK5-branes \cite{Kimura:2014wga}, or KK-vortices \cite{Okada:2014wma}.
A $[1,0]^T_{k5}$-brane is a KK-vortex and a $[0,1]^T_{k5}$-brane is its $SL(2,{\mathbb Z})_{\tau}$ dual object.

\item $[p,q]^E_{fn}$-brane ($n = 0,1$). 
This is a defect brane in $(n+3)$ dimensions.
More clearly, a $[1,0]^E_{f1}$-brane is an F-string and a $[0,1]^E_{f1}$-brane is its $SL(2,{\mathbb Z})_E$ dual, i.e., a $1^6_4$-brane. 
In the same way, a $[1,0]^E_{f0}$-brane is a pp-wave symbolized as P in Figure \ref{fig:E-dualities}, and a $[0,1]^E_{f0}$-brane is a $0^{(1,6)}_4$-brane.

\end{itemize}
In addition, we also use the notation introduced in \cite{deBoer:2010ud, deBoer:2012ma}.
For instance, a D7-brane extending in the 1234567-directions is referred to as D7(1234567)-brane, and an exotic $5^2_3$-brane extending in the 12345-direction with two isometries along the 67-directions as $5^2_3$(12345,67)-brane.

\vspace{5mm}

The structure of this paper is as follows.
In section \ref{sect:F-theory},
we briefly review the role of seven-branes in F-theory.
We argue the monodromy and the branch cut, and branes crossing the cut.
We also discuss a configuration given by the web of five-branes with multiple seven-branes and its superconformal limit.
In section \ref{sect:BJ-ts5},
we apply the string dualities to the brane junctions in F-theory.
We find the object ending on the exotic $5^2_2$-brane.
Furthermore, we also find three new brane junctions originated from the $[p,q]^T_{s5}$-brane.
In section \ref{sect:Web-SCFT},
we construct three webs of branes in the presence of multiple exotic five-branes.
Analogous to the original five-branes web with seven-branes, we qualitatively obtain $SU(2)$ gauge theories with $n$ flavors and their superconformal limit with enhanced $E_{n+1}$ symmetry in five, four, and three dimensions.
Section \ref{sect:summary} is devoted to the summary and discussions.
In appendix \ref{app:ex-EBJ}, we demonstrate an example of the exotic brane junctions which is invariant under the string S-duality.

\section{Short review of seven-branes in F-theory}
\label{sect:F-theory}

In this section we review the property of seven-branes in F-theory.
In section \ref{sect:single7}, we focus on a single $[p,q]^S_7$(1234567)-brane and its branch cut in its transverse 89-plane.
When a $(r,s)$-string and a $(r,s)_5$-brane cross the branch cut, they are transformed by the monodromy of the $[p,q]^S_7$-brane \cite{Gaberdiel:1997ud, DeWolfe:1998zf, DeWolfe:1998yf, DeWolfe:1998eu, DeWolfe:1998pr}.
In section \ref{sect:multi7}, we review the web of five-branes with multiple $[p,q]^S_7$-branes. This is the five-dimensional version of the Hanany-Witten brane configuration \cite{Hanany:1996ie} and has been developed by \cite{Aharony:1997bh, DeWolfe:1999hj, Benini:2009gi, Bergman:2013ala, Bao:2013pwa, Hayashi:2013qwa, Kim:2014nqa, Kim:2015jba}.

\subsection{Brane junctions in F-theory}
\label{sect:single7}

First of all, one introduces a complex variable $\rho = C^{(0)} + \I \e^{-\phi}$, where $C^{(0)}$ and $\phi$ are the axion and the dilaton, respectively.
$\rho$ is called the complex modulus which depends only on the coordinate $z \equiv x^8 + \I x^9$.
In ten dimensions, $\rho$ couples to a seven-brane magnetically.
In addition, in the equations of motion in type IIB string theory, this couples to two three-forms $F^i_{(3)} = (F_{(3)}, H_{(3)})^{\T}$, where $F_{(3)} = \d C_{(2)}$ and $H_{(3)} = \d B_{(2)}$ are the R-R and the NS-NS three-form field strengths, respectively.
Under the $SL(2,{\mathbb Z})_S$ S-duality, $F^i_{(3)}$ is transformed as a doublet:
\bsubeq
\begin{gather}
\rho \ \to \ \frac{a \rho + b}{c \rho + d}
\, , \ls
\Lambda^i{}_j \ \equiv \ \left( 
{\renewcommand{\arraystretch}{.85}
\begin{array}{cc}
a & b
\\
c & d
\end{array}
}
\right)
\ \in \ SL(2,{\mathbb Z})_S
\, , \\
F^i_{(3)} \ \to \ 
\Lambda^i{}_j \, F^j_{(3)}
\, . 
\end{gather}
\esubeq
This implies that a configuration of a single F-string, given as the $(1,0)$-string, is rotated to a $(p,q)$-string under the $SL(2,{\mathbb Z})$ duality group in such a way that
\begin{align}
\left(
{\renewcommand{\arraystretch}{.85}
\begin{array}{c}
1
\\
0
\end{array}
}
\right)
\ &\to \ 
\left(
{\renewcommand{\arraystretch}{.85}
\begin{array}{c}
p
\\
q
\end{array}
}
\right)
\ = \ 
g_{p,q} \left(
{\renewcommand{\arraystretch}{.85}
\begin{array}{c}
1 
\\
0
\end{array}
}
\right)
\, , \ls
g_{p,q}
\ = \ 
\left(
{\renewcommand{\arraystretch}{.85}
\begin{array}{cc}
p & r 
\\
q & s
\end{array}
}
\right) \ \in \ SL(2,{\mathbb Z})
\, . \label{10-pq-string}
\end{align} 
Here the integers $r$ and $s$ are arbitrary as far as they satisfy $ps - qr = 1$.
Because $F^i_{(3)}$ is a doublet of the $SL(2,{\mathbb Z})_S$ duality,
the $(p,q)$-string is a bound state of $p$ F-strings and $q$ D-strings.

A D7-brane is a seven-brane on which an F-string is ending.
Performing the duality transformation (\ref{10-pq-string}), 
the D7-brane is also transformed to a $[p,q]^S_7$-brane on which a $(p,q)$-string is ending.
Consider a simple situation that a D7-brane (i.e., a $[1,0]^S_7$-brane) is dualized to a $[0,1]^S_7$-brane.
This seven-brane is nothing but the exotic $7_3$-brane on which a D-string is ending.
We should keep in mind this phenomenon for later discussions.

One considers the monodromy of a single $[p,q]^S_7$(1234567)-brane sitting on the origin of the 89-plane.
Since this is an extended object of codimension two in ten-dimensional type IIB string theory,
the $SL(2,{\mathbb Z})_S$ S-duality branch cut appears in the 89-plane.
It is known that the monodromy of a single $[p,q]^S_7$-brane is given by the S-duality transformation of that of a single $[1,0]^S_7$-brane in the following way.
Going around the $[1,0]^S_7$-brane counterclockwise as $z \to z \, \e^{2 \pi \I}$, 
the complex modulus $\rho$ is transformed as follows:
\begin{align}
\rho \ &\to \ 
\frac{a \rho + b}{c \rho + d}
\ = \ 
\rho + 1
\, , \ls
\left(
{\renewcommand{\arraystretch}{.85}
\begin{array}{cc}
a & b
\\
c & d
\end{array}
}
\right)
\ \in \ SL(2,{\mathbb Z})
\, . 
\end{align}
Then the transformation matrix $M_{[1,0]^S_7}$ and its inverse $K_{[1,0]^S_7}$ is obtained as
\begin{align}
M_{[1,0]^S_7} \ &\equiv \ 
\left(
{\renewcommand{\arraystretch}{.85}
\begin{array}{cc}
1 & 1 
\\
0 & 1
\end{array}
}
\right)
\, , \ls
K_{[1,0]^S_7} \ \equiv \ 
(M_{[1,0]^S_7})^{-1}
\ = \ 
\left(
{\renewcommand{\arraystretch}{.85}
\begin{array}{cc}
1 & -1 
\\
0 & 1
\end{array}
}
\right)
\, . \label{10s7-monodromy}
\end{align}
Since a $[p,q]^S_7$-brane originates from a D7-brane via the S-duality (\ref{10-pq-string}), the matrix is also transformed in terms of $g_{p,q}$:
\begin{align}
K_{[p,q]^S_7} \ &\equiv \ 
\big( g_{p,q} \, M_{[p,q]^S_7} \, g_{p,q}^{-1} \big)^{-1}
\ = \ 
\left(
{\renewcommand{\arraystretch}{.85}
\begin{array}{cc}
1 + pq & -p^2 
\\
q^2 & 1 - pq
\end{array}
}
\right)
\, . \label{Kpqs7-monodromy}
\end{align}
Since $K_{[p,q]^S_7}$ is quite useful when we consider a certain object crosses the branch cut of the $[p,q]^S_7$-brane counterclockwise.
Hence this is referred to as the monodromy matrix.
Here a configuration that an $(r,s)$-string crosses the branch cut of the $[p,q]^S_7$-brane is illustrated in Figure \ref{fig:pqs7-rss1}:
\begin{center}
\slb{.85}{\includegraphics[bb=0 0 159 95]{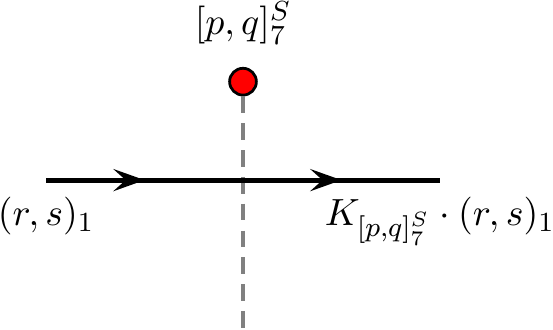}}
\figcaption{A $[p,q]^S_7$(1234567)-brane as a point in the 89-plane and an $(r,s)$-string along the 8th-direction crossing the branch cut (the vertical dashed line) from the $[p,q]^S_7$-brane. 
This $(r,s)$-string is transformed by the monodromy matrix $K_{[p,q]^S_7}$ when it crosses the branch cut from the left.}
\label{fig:pqs7-rss1}
\end{center}
Because of the charge conservation, the Hanany-Witten transition \cite{Hanany:1996ie} occurs as in Figure \ref{fig:pqs7-rss1-HW}:
\begin{center}
\slb{.85}{\includegraphics[bb=0 0 159 113]{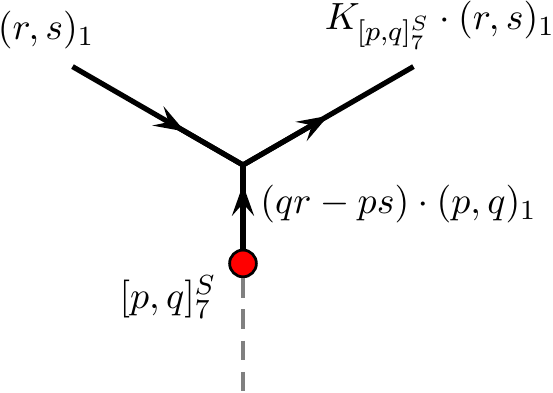}}
\figcaption{The Hanany-Witten transition of Figure \ref{fig:pqs7-rss1} occurs when the $[p,q]^S_7$-brane crosses the $(r,s)$-string. Here $qr - ps = \pm 1$ is required by the charge conservation.}
\label{fig:pqs7-rss1-HW}
\end{center}
Here three strings gather at a single point. This is called a string junction. The charge of the string from the $[p,q]^S_7$-brane is determined by the monodromy matrix:
\begin{align}
K_{[p,q]^S_7} 
\left(
{\renewcommand{\arraystretch}{.85}
\begin{array}{c}
r
\\
s
\end{array}
}
\right)
\ &= \ 
\left(
{\renewcommand{\arraystretch}{.85}
\begin{array}{cc}
1 + pq & -p^2 
\\
q^2 & 1 - pq
\end{array}
}
\right)
\left(
{\renewcommand{\arraystretch}{.85}
\begin{array}{c}
r
\\
s
\end{array}
}
\right)
\ = \ 
\left(
{\renewcommand{\arraystretch}{.85}
\begin{array}{c}
r
\\
s
\end{array}
}
\right)
+ (qr - ps) 
\left(
{\renewcommand{\arraystretch}{.85}
\begin{array}{c}
p
\\
q
\end{array}
}
\right)
\, . \label{charge-pq-rs}
\end{align}
It is noticed that the factor $qr -ps$ should be $\pm 1$.
The negative sign is also allowed since this implies that a $(-p,-q)$-string whose orientation is opposite to that of the $(p,q)$-string also ends on the $[p,q]^S_7$-brane. 

In a similar fashion, one can also consider a configuration that an $(r,s)_5$-brane crosses the branch cut of the $[p,q]^S_7$-brane as in Figure \ref{fig:pqs7-rss5}:
\begin{center}
\slb{.68}{\includegraphics[bb=0 0 159 95]{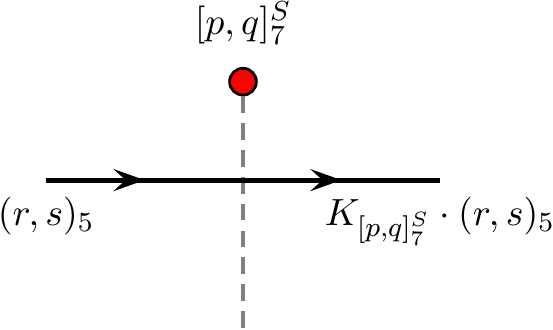}}
\raisebox{12mm}{\slb{.85}{$\xrightarrow{\text{HW}}$}}  
\slb{.68}{\includegraphics[bb=0 0 159 113]{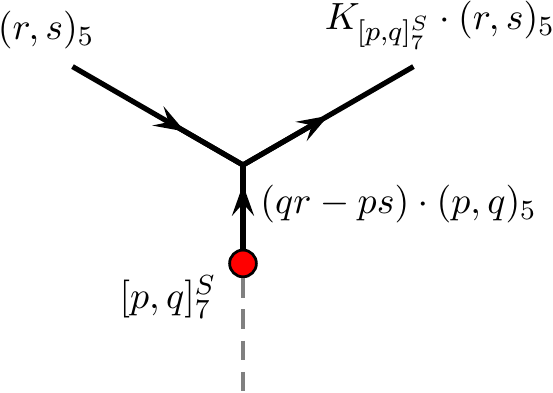}}
\raisebox{13mm}{
\slb{.68}{
\renewcommand{\arraystretch}{1.1}
\begin{tabular}{c:c||c|cccc|ccc|c:c} \hline
\multicolumn{2}{c||}{IIB} & 0 & 1 & 2 & 3 & 4 & 5 & 6 & 7 & 8 & 9
\\ \hline\hline
\multirow{2}{*}{$[p,q]^S_{7}$} 
& $p$ D7 & $-$ & $-$ & $-$ & $-$ & $-$ & $-$ & $-$ & $-$ & & 
\\
& $q$ $7_3$ & $-$ & $-$ & $-$ & $-$ & $-$ & $-$ & $-$ & $-$ & & 
\\ \hline
& D5 & $-$ & $-$ & $-$ & $-$ & $-$ & & & & $-$ &
\\
& NS5 & $-$ & $-$ & $-$ & $-$ & $-$ & & & & & $-$ 
\\ \hline
\multirow{2}{*}{$(r,s)_5$} 
& $r$ D5 & $-$ & $-$ & $-$ & $-$ & $-$ & & & & \multicolumn{2}{c}{\multirow{2}{*}{angle}} 
\\
& $s$ NS5 & $-$ & $-$ & $-$ & $-$ & $-$ & & & & \multicolumn{2}{c}{} 
\\ \hline
\end{tabular}
}
}
\figcaption{A $[p,q]^S_7$(1234567)-brane as a point in the 89-plane and an $(r,s)_5$-brane crossing the branch cut, and its Hanany-Witten transition ``HW''.}
\label{fig:pqs7-rss5}
\end{center}
Here, in order to preserve the 1/4-BPS condition, we assume that a $(1,0)_5$-brane, i.e., a D5-brane extends in the 12348-directions, while a $(0,1)_5$-brane as an NS5-brane extends in the 12349-directions.
In this setup, the $(r,s)_5$-brane extends in the 1234X-directions, where X means a certain direction in the 89-plane with angle. 
The angle is determined by the co-prime charges $r$ and $s$.
It is remarked that the $(p,q)_5$-brane ends on the $[p,q]^S_7$-brane because a $(p,q)$-string ends on both the $(p,q)_5$-brane and the $[p,q]^S_7$-brane.
The five-brane charges also satisfy the equation (\ref{charge-pq-rs}).

\subsection{Web of five-branes with multiple seven-branes}
\label{sect:multi7}

In terms of the five-brane junctions in Figure \ref{fig:pqs7-rss5}, 
one can construct a web of five-branes on which a five-dimensional supersymmetric gauge theory with flavors emerges.
In particular, thanks to the technology developed by the recent works \cite{Benini:2009gi, Bergman:2013ala, Bao:2013pwa, Hayashi:2013qwa, Kim:2014nqa, Kim:2015jba},
one can analyze the strong gauge coupling limit.
First, three seven-branes are introduced as follows:
\bsubeq \label{ABC-branes}
\begin{alignat}{2}
\text{\A-brane} \ &= \ [1,0]^S_7 
\, , &\ls
K_A \ &= \ 
\left(
{\renewcommand{\arraystretch}{.85}
\begin{array}{cc}
1 & -1 
\\
0 & 1
\end{array}
}
\right)
\, , \label{A-brane} \\
\text{\B-brane} \ &= \ [1,-1]^S_7 
\, , &\ls
K_B \ &= \ 
\left(
{\renewcommand{\arraystretch}{.85}
\begin{array}{cc}
0 & -1 
\\
1 & 2
\end{array}
}
\right)
\, , \label{B-brane} \\
\text{\C-brane} \ &= \ [1,1]^S_7 
\, , &\ls
K_C \ &= \ 
\left(
{\renewcommand{\arraystretch}{.85}
\begin{array}{cc}
2 & -1 
\\
1 & 0
\end{array}
}
\right)
\, . \label{C-brane}
\end{alignat}
\esubeq
Here $K_{A,B,C}$ are the monodromy matrices of the $\A$-, $\B$- and $\C$-branes, respectively.

Consider a brane configuration on which the five-dimensional $\N=1$ $SU(2)$ gauge theory with $n$ flavors is realized as in Figure \ref{fig:web5-7}:
\begin{center}
\slb{.8}{\includegraphics[bb=0 0 143 143]{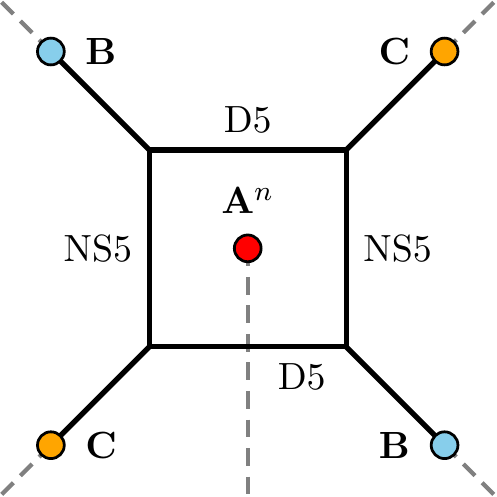}}
\LS
\raisebox{20mm}{
\slb{.8}{
\renewcommand{\arraystretch}{1.1}
\begin{tabular}{c:c||c|cccc|ccc|c:c} \hline
\multicolumn{2}{c||}{IIB} & 0 & 1 & 2 & 3 & 4 & 5 & 6 & 7 & 8 & 9
\\ \hline\hline
\multirow{2}{*}{$\A^n$} 
& $n$ D7 & $-$ & $-$ & $-$ & $-$ & $-$ & $-$ & $-$ & $-$ & & 
\\
& 0 $7_3$ & $-$ & $-$ & $-$ & $-$ & $-$ & $-$ & $-$ & $-$ & & 
\\ \hline
& D5 & $-$ & $-$ & $-$ & $-$ & $-$ & & & & $-$ &
\\
& NS5 & $-$ & $-$ & $-$ & $-$ & $-$ & & & & & $-$
\\ \hline
\multirow{2}{*}{$(r,s)_5$} 
& $r$ D5 & $-$ & $-$ & $-$ & $-$ & $-$ & & & & \multicolumn{2}{c}{\multirow{2}{*}{angle}} 
\\
& $s$ NS5 & $-$ & $-$ & $-$ & $-$ & $-$ & & & & \multicolumn{2}{c}{} 
\\ \hline
\end{tabular}
}
}
\figcaption{A five-brane quadrilateral with $n$ coincident $\A$-branes, two $\B$-branes and two $\C$-branes. 
The table expresses the extending directions of the $\A^n$-branes and the five-branes.
In the gravity decoupling limit, this system describes the 5D $\N=1$ $SU(2)$ gauge theory with $n$ flavors along the 01234-directions on the horizontal D5-branes.}
\label{fig:web5-7}
\end{center}
Here two $(1,-1)_5$-brane and two $(1,-1)_5$-branes are ending on $\B$-branes and $\C$-branes, respectively.
It is noticed that one can freely move both the $\B$-branes and the $\C$-branes along the geodesics, i.e., the $(1,-1)$-direction and the $(1,1)$-direction in Figure \ref{fig:web5-7}, respectively.
We note that the $(1,0)$-direction and the $(0,1)$-direction in the figure correspond to the 8th-direction and the 9th-direction in spacetime, respectively.
In the gravity decoupling limit, one can trace the five-dimensional $\N=1$ $SU(2)$ gauge theory on the horizontal D5-branes suspended between two vertical NS5-branes. 
Now the finite distance of two NS5-branes is set to $L$.
This is a five-dimensional version of the Hanany-Witten setup \cite{Hanany:1996ie}.

For simplicity, one restricts the number of $n = 5,6,7$\footnote{One can analyze gauge theories with $0 \leq n \leq 8$ to find their healthy superconformal limit, while in the $n \geq 9$ case there are no fixed points \cite{Kim:2015jba}.}.
One can freely move the $\B$-branes and the $\C$-branes into the five-brane quadrilateral without changing the worldvolume theory.
Once all the $\B$- and $\C$-branes go into the quadrilateral, the $(1,-1)_5$-branes and the $(1,1)_5$-branes are annihilated by the Hanany-Witten transitions.
Further, the quadrilateral is highly curved and becomes a loop by the back reactions from the seven-branes. Hence, without crossing the branch cuts,
one finds that the seven-branes are arrayed as $\A^n \B\C\B\C$.
Now, move the monodromy branch cuts and reorder the seven-branes in such a way as $\A^n \B \C \C \X_{[3,1]}$, where $\X_{[3,1]}$ is a $[3,1]^S_7$-brane.
In the small size limit $L \to 0$ where the five-brane loop is shrunk,
the $\A^n\B\C\C$-branes are collapsed to a point and are converted to 
$\E_{n+1}$ inside the loop, while $\X_{[3,1]}$ is gone infinitely far away from the loop.
In this limit, the gauge coupling constant on the five-brane becomes strong and  the gauge theory goes to a non-trivial UV fixed point.
Eventually, the five-dimensional worldvolume remains along the 01234-directions.
There, an $\N=1$ superconformal field theory with enhanced $E_{n+1}$ symmetry \cite{Seiberg:1996bd} appears \cite{DeWolfe:1999hj}.
This is illustrated in Figure \ref{fig:F-HW-small-En}:
\begin{center}
\slb{.8}{\includegraphics[bb=0 0 143 143]{./Anbox-F.pdf}}
\raisebox{18mm}{$\xrightarrow{\text{HW}}$} 
\raisebox{10mm}{\slb{.8}{\includegraphics[bb=0 0 97 69]{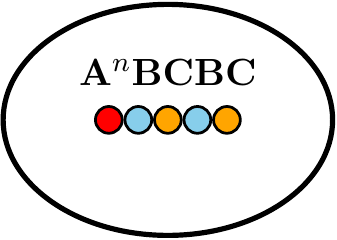}}}
\raisebox{18mm}{$\xrightarrow{\text{reordering}}$} 
\raisebox{10mm}{\slb{.8}{\includegraphics[bb=0 0 97 69]{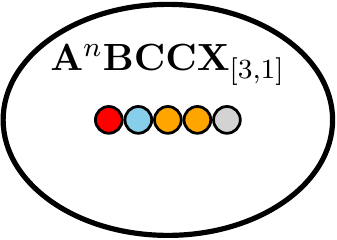}}}
\raisebox{18mm}{$\xrightarrow{\text{small}}$} 
\raisebox{16mm}{\slb{.8}{\includegraphics[bb=0 0 77 37]{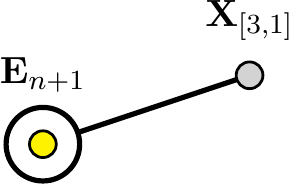}}}
\figcaption{After the Hanany-Witten transitions of Figure \ref{fig:web5-7}, one reorders the seven-branes by moving their branch cuts. Further, one takes the small size limit of the five-brane loop when $n = 5,6,7$.}
\label{fig:F-HW-small-En}
\end{center}

\section{Brane junctions by single exotic five-brane}
\label{sect:BJ-ts5}

In this section, we study the brane junctions by exotic five-branes associated with those in F-theory.
Performing the string dualities, we will reveal the property of the exotic $5^2_2$-brane\footnote{There is also an interesting approach to understand the monodromies of exotic five-branes or T-folds \cite{Lust:2015yia}.}.

\subsection{Monodromy matrix}
\label{sect:ts5-mono}

We begin with a single D7-brane in ten dimensions.
Applying the T-dualities along the 67-directions to the D7(1234567)-brane, 
we obtain a D5(12345)-brane.
We assume that the compactified 67-directions become a two-torus $T^2_{67}$ and smeared.
Then the complex modulus $\rho(z)$ of the D5-brane is now given as $\rho(z) = C^{(2)}_{67} + \I \e^{- 2 \phi}$ \cite{Sakatani:2014hba}.
When we take the S-duality, the D5(12345)-brane becomes a smeared NS5(12345)-brane with the complex modulus $\rho(z) = B^{(2)}_{67} + \I \e^{2 \phi} = B^{(2)}_{67} + \I \sqrt{\det G_{mn}}$, where $G_{mn}$ is the metric on the two-torus $T^2_{67}$.
On the other hand,
if we take the S-duality to the original D7(1234567)-brane in ten dimensions,
we find that a $[0,1]^S_7$(1234567)-brane (i.e., an exotic $7_3$(1234567)-brane) appears. 
Compactifying the 67-directions to the two-torus and performing the T-dualities along the 67-directions to the $7_3$(1234567)-brane, 
we find an exotic $5^2_3$(12345,67)-brane in Figure \ref{fig:E-dualities}.
Now, when we further apply the S-duality to the $5^2_3$(12345,67)-brane, 
an exotic $5^2_2$(12345,67)-brane appears.
This exotic brane is an $SL(2,{\mathbb Z})_{\rho} \subset SO(2,2;{\mathbb Z})_T$ dual to the defect NS5(12345)-brane. 
Hence, if we take the zero size limit of the two torus $T^2_{67}$,
the NS5-brane from the D7-brane via the ST$_{67}$-dualities is exotic dual to the $5^2_2$(12345,67)-brane from the $7_3$-brane via the ST$_{67}$-dualities.
This implies that the $[p,q]^S_7$-brane in ten-dimensional theory becomes a $[p,q]^T_{s5}$-brane of codimension two in {\it eight}-dimensional theory.
As mentioned before, this is nothing but the ``defect $(p,q)$ five-brane'' discussed in \cite{Kimura:2014wga}.

The single $[p,q]^T_{s5}$-brane also has a non-trivial monodromy originated from the $SL(2,{\mathbb Z})_{\rho}$ subgroup of the T-duality group $SO(2,2;{\mathbb Z})_T$ acting on the compactified two-torus $T^2_{67}$.
Indeed, the original $SL(2,{\mathbb Z})_S$ monodromy of the $[p,q]^S_7$-brane in ten dimensions is mapped to the $SL(2,{\mathbb Z})_{\rho}$ monodromy via the ST$_{67}$-dualities.
Hence the monodromy matrix of the single $[p,q]^T_{s5}$-brane is exactly the same as that of the single $[p,q]^S_7$-brane:
\begin{align}
K_{[p,q]^T_{s5}} \ &= \
K_{[p,q]^S_7} \ = \ 
\left(
{\renewcommand{\arraystretch}{0.85}
\begin{array}{cc}
1 + pq & - p^2
\\
q^2 & 1 - pq
\end{array}
}
\right)
\, . \label{Kpqts5-monodromy}
\end{align}
In addition, following the same procedure,
we can also immediately obtain the monodromy matrices of the other single $[p,q]$-branes introduced in section \ref{sect:introduction}:
Applying the T$_{(7-n)}$-dualities along the worldvolume directions of the $[p,q]^S_7$-brane\footnote{We express the T-dualities along $n$ individual directions as the T$_{(n)}$-dualities, for short.}, 
the monodromy matrix of the single $[p,q]^E_{dn}$-brane is obtained as $K_{[p,q]^E_{dn}} = K_{[p,q]^S_7}$, where $0 \leq n \leq 7$.
Furthermore, the monodromy matrices of the single $[p,q]^E_{f1}$-brane (i.e., the pair of F-strings and $1^6_4$-branes) and the single $[p,q]^E_{f0}$-brane (the pair of pp-waves and $0^{(1,6)}_4$-branes) are also given as $K_{[p,q]^E_{f1}} = K_{[p,q]^S_7} = K_{[p,q]^E_{f0}}$.

As discussed above, we can borrow the techniques on the single $[p,q]^S_7$-brane in ten dimensions in order to consider configurations of the single $[p,q]^T_{s5}$-brane. 
This is also true when we consider objects sensitive to the $SL(2,{\mathbb Z})_{\rho}$ monodromy branch cut originated from the single $[p,q]^T_{s5}$-brane, and configurations of multiple branes with multiple $[p,q]^T_{s5}$-branes.

We have to argue which object is ending on the $[p,q]^T_{s5}$-brane.
Recall that the $(p,q)$-string ending on the $[p,q]^S_7$-brane is a bound state of $p$ F-strings and $q$ D-strings.
Applying the ST$_{67}$-dualities to the $[p,q]^S_7$(1234567)-brane,
we find that a bound state of $p$ D-strings and $q$ D3-branes wrapped on the two-torus $T^2_{67}$.
If the size of $T^2_{67}$ is very small, these $q$ wrapped D3-branes can be seen as $q$ ``D-strings'' in eight dimensions.
Then we see that the two kinds of D-strings are ending on the $[p,q]^T_{s5}$-brane.
In other words,
a $[1,0]^T_{s5}$-brane is a defect NS5(12345)-brane on which a D-string is ending,
while a $[0,1]^T_{s5}$-brane is an exotic $5^2_2$(12345,67)-brane on which a D3-brane wrapped on $T^2_{67}$ is ending.
This situation is consistent with the relation between the defect NS5(12345)-brane and the $5^2_2$(12345,67)-brane under the T$_{67}$-dualities.
Hence we now understand that the (small) excitations on the worldvolume of the exotic $5^2_2$-brane is governed by the oscillation modes of the D3-brane wrapped on the isometry directions.

We can further discuss the ``$(p,q)$-strings'' ending on the $[p,q]$-branes in diverse dimensions.
Since the $[p,q]^E_{dn}$-brane is obtained from the $[p,q]^S_7$-brane by the T$_{(n)}$-dualities,
the object ending on the $[p,q]^E_{dn}$-brane is a bound state of $p$ F-strings and $q$ D$(n+1)$-branes wrapped on the compact $n$-torus $T^n$.
Furthermore, the single $[p,q]^E_{f1}$-brane is a descendant of the $[p,q]^S_7$-brane by the ST$_{(6)}$-dualities, the object ending on it is a bound state of $p$ D-strings and $q$ $7_3$-branes wrapped on the compact six-torus $T^6$.
The object ending on the $[p,q]^E_{f0}$-brane is a bound state of $p$ D2-branes wrapped on $S^1$ and $q$ $7^{(1,0)}_3$-branes wrapped on the compact seven-torus $T^7$.

\subsection{String junction}

First, we consider the ST$_{67}$-dualities of the configurations in Figures \ref{fig:pqs7-rss1} and \ref{fig:pqs7-rss1-HW}.
As mentioned above, 
the $[p,q]^S_7$-brane with its branch cut is mapped to the $[p,q]^T_{s5}$-brane with the corresponding branch cut in each figure.
The monodromy matrix is also mapped as (\ref{Kpqts5-monodromy}).
On the other hand, the $(r,s)$-string, i.e., the bound state of $r$ F-strings and $s$ D-strings in Figures \ref{fig:pqs7-rss1} and \ref{fig:pqs7-rss1-HW} is transformed to $r$ D-strings and $s$ D3-branes wrapped on the compact two-torus $T^2_{67}$ via the ST$_{67}$-dualities.
If the size of $T^2_{67}$ is infinitesimal, the wrapped D3-branes behave as D-strings in eight dimensions.
Hence we again interpret that the bound state of $r$ D-strings and $s$ wrapped D3-branes as the ``$(r,s)$-string''.
This new $(r,s)$-string is sensitive to the branch cut from the $[p,q]^T_{s5}$-brane as in the original configuration.
Then, via the Hanany-Witten transition, the $(r,s)$-string obtain a three string junction as far as the charge conservation (\ref{charge-pq-rs}) is satisfied.
The new configuration, i.e., the $(r,s)$-string crossing the $[p,q]^T_{s5}$-brane monodromy branch cut and its Hanany-Witten transition, is illustrated in Figure \ref{fig:pqts5-rss1}:
\begin{center}
\slb{.68}{\includegraphics[bb=0 0 160 95]{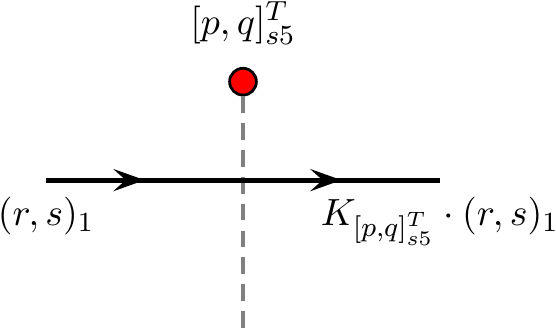}}
\raisebox{12mm}{\slb{.85}{$\xrightarrow{\text{HW}}$}}
\slb{.68}{\includegraphics[bb=0 0 160 113]{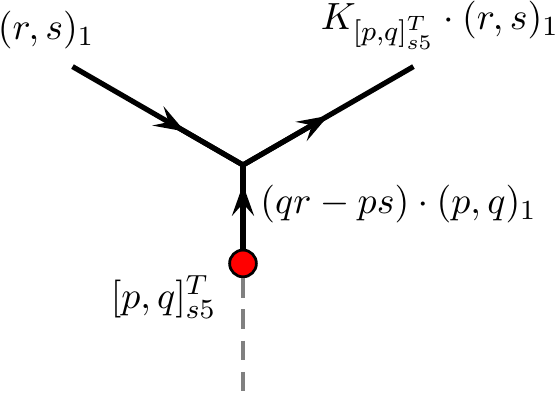}}
\figcaption{A $[p,q]^T_{s5}$-brane and an $(r,s)$-string crossing the branch cut, and its Hanany-Witten transition. This is, for instance, ST$_{67}$-dual of Figure \ref{fig:pqs7-rss1}.}
\label{fig:pqts5-rss1}
\end{center}
Let us focus on the exotic $5^2_2$-brane, i.e., the $[0,1]^T_{s5}$-brane in Figure \ref{fig:pqts5-rss1}.
We immediately find that the $(r,s)$-string crossing the branch cut is transformed to the $(r,r+s)$-string. 
This implies that the D-string is sensitive to the monodromy of the $5^2_2$-brane.
Furthermore, the wrapped D3-brane is created from the $5^2_2$-brane via the Hanany-Witten transition.

\subsection{Brane junctions}

Next, we argue the five-brane junctions originated from the single $[p,q]^T_{s5}$-brane.
Different from the string junction in Figure \ref{fig:pqts5-rss1}, 
we find three distinct configurations from the original one in Figure \ref{fig:pqs7-rss5}. 
This is because of the difference of T-dualized directions.

The first configuration is given by the ST$_{67}$-dualities of Figure \ref{fig:pqs7-rss5}.
Since the original $(r,s)_5$-brane extends in the 1234X-direction where X is a certain direction in the 89-plane,
this five-brane is transformed to a bound state of $r$ exotic $7_3$(1234678)-branes and $s$ exotic $5^2_3$(12349,67)-branes.
Because of the charge conservation, this bound state has a certain angle in the 89-plane depending on the co-prime charges $r$ and $s$.
Analogous to the original configuration, the new $(r,s)_5$-brane is also sensitive to the $[p,q]^T_{s5}$-brane monodromy.
If the $[p,q]^T_{s5}$-brane crosses the $(r,s)_5$-brane, the Hanany-Witten transition occurs if $qr - ps = \pm 1$.
This configuration is illustrated in Figure \ref{fig:pqts5-rss5}:
\begin{center}
\slb{.68}{\includegraphics[bb=0 0 160 95]{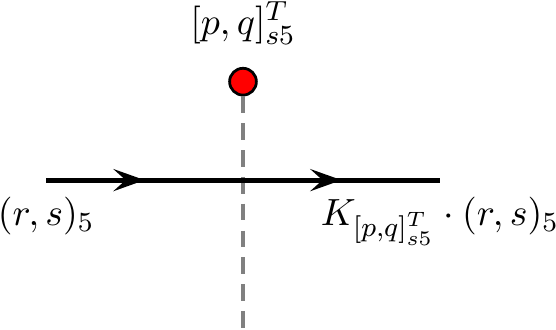}}
\raisebox{12mm}{\slb{.85}{$\xrightarrow{\text{HW}}$}} 
\slb{.68}{\includegraphics[bb=0 0 160 113]{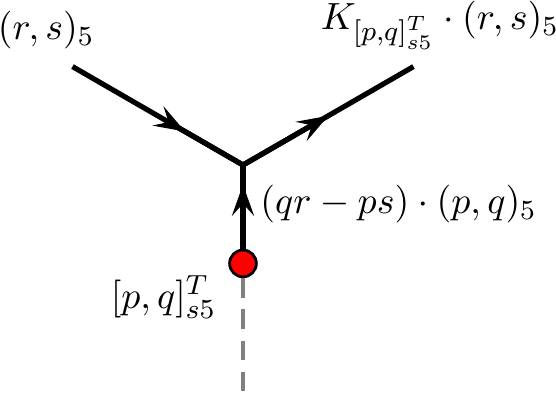}}
\raisebox{13mm}{
\slb{.68}{
\renewcommand{\arraystretch}{1.1}
\begin{tabular}{c:c||c|cccc|c|cc|c:c} \hline
\multicolumn{2}{c||}{IIB} & 0 & 1 & 2 & 3 & 4 & 5 & \txc{6} & \txc{7} & 8 & 9
\\ \hline\hline
\multirow{2}{*}{$[p,q]^T_{s5}$} 
& $p$ NS5 & $-$ & $-$ & $-$ & $-$ & $-$ & $-$ & & & & 
\\
& $q$ $5^2_2$ & $-$ & $-$ & $-$ & $-$ & $-$ & $-$ & \tcc{2} & \tcc{2} & & 
\\ \hline
& ``$7_3$'' & $-$ & $-$ & $-$ & $-$ & $-$ & & $-$ & $-$ & $-$ &
\\
& $5^2_3$ & $-$ & $-$ & $-$ & $-$ & $-$ & & \tcc{2} & \tcc{2} & & $-$ 
\\ \hline
\multirow{2}{*}{$(r,s)_{5}$} 
& $r$ ``$7_3$'' & $-$ & $-$ & $-$ & $-$ & $-$ & & $-$ & $-$ & \multicolumn{2}{c}{\multirow{2}{*}{angle}} 
\\
& $s$ $5^2_3$ & $-$ & $-$ & $-$ & $-$ & $-$ & & \tcc{2} & \tcc{2} & \multicolumn{2}{c}{} 
\\ \hline
\end{tabular}
}}
\figcaption{A $[p,q]^T_{s5}$-brane and an $(r,s)_5$-brane crossing the branch cut, and its Hanany-Witten transition. This is ST$_{67}$-dual of Figure \ref{fig:pqs7-rss5}.
Here the symbol ``\tcc{2}'' implies that the mass of the exotic brane depends on the square of the compact radius of the corresponding direction \cite{Kimura:2016xzd}.
The terminology ``$7_3$'' denotes that the $7_3$-brane is wrapped on the compact 67-directions.}
\label{fig:pqts5-rss5}
\end{center}
This configuration preserves one quarter supersymmetries. This is guaranteed by the projection conditions of the supersymmetry on the standard branes and the exotic branes \cite{LozanoTellechea:2000mc, deBoer:2012ma, Kimura:2016xzd}.
We can find the following properties of the $5^2_2$-brane when $[p,q] = [0,1]$ in Figure \ref{fig:pqts5-rss5}:
A wrapped $7_3$-brane is sensitive to the branch cut.
This is transformed to a bound state of a wrapped $7_3$-brane and a $5^2_3$-brane.
This implies that the $5^2_3$-brane is created from the $5^2_2$-brane by the Hanany-Witten transition.

The second configuration is obtained by the ST$_{47}$-dualities of Figure \ref{fig:pqs7-rss5}.
In this case, the original $(r,s)_5$-brane becomes a bound state of 
$r$ NS5-branes wrapped on the compact circle $S^1_7$ along the 7th-direction and $s$ KK5-branes wrapped on the same compact $S^1_7$.
If the compact circles along the 47-directions are infinitesimal, the five-branes can be seen as four-branes.
Then we rename the new object crossing the branch cut as the $(r,s)_4$-brane.
This new object is also transformed by the $[p,q]^T_{s5}$-brane monodromy.
The configuration is illustrated in Figure \ref{fig:pqts5-rss4}:
\begin{center}
\slb{.68}{\includegraphics[bb=0 0 161 95]{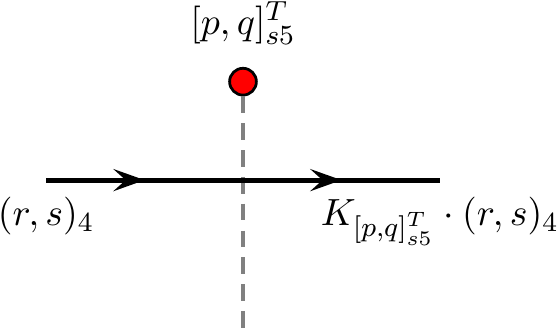}}
\raisebox{12mm}{\slb{.85}{$\xrightarrow{\text{HW}}$}} 
\slb{.68}{\includegraphics[bb=0 0 161 113]{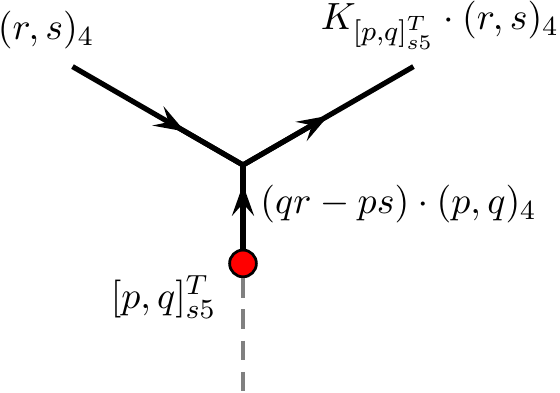}}
\raisebox{13mm}{
\slb{.68}{
\renewcommand{\arraystretch}{1.1}
\begin{tabular}{c:c||c|ccc|c|cc|c|c:c} \hline
\multicolumn{2}{c||}{IIB} & 0 & 1 & 2 & 3 & \txc{4} & 5 & 6 & \txc{7} & 8 & 9
\\ \hline\hline
\multirow{2}{*}{$[p,q]^T_{s5}$} 
& $p$ NS5 & $-$ & $-$ & $-$ & $-$ & & $-$ & $-$ & & & 
\\
& $q$ $5^2_2$ & $-$ & $-$ & $-$ & $-$ & \tcc{2} & $-$ & $-$ & \tcc{2} & & 
\\ \hline
& ``NS5'' & $-$ & $-$ & $-$ & $-$ & & & & $-$ & $-$ &
\\
& ``KK5'' & $-$ & $-$ & $-$ & $-$ & $-$ & & & \tcc{2} & & $-$ 
\\ \hline
\multirow{2}{*}{$(r,s)_4$} 
& $r$ ``NS5'' & $-$ & $-$ & $-$ & $-$ & & & & $-$ & \multicolumn{2}{c}{\multirow{2}{*}{angle}} 
\\
& $s$ ``KK5'' & $-$ & $-$ & $-$ & $-$ & $-$ & & & \tcc{2} & \multicolumn{2}{c}{} 
\\ \hline
\end{tabular}
}}
\figcaption{A $[p,q]^T_{s5}$-brane and an $(r,s)_4$-brane crossing the branch cut, and its Hanany-Witten transition. This is ST$_{47}$-dual of Figure \ref{fig:pqs7-rss5}.}
\label{fig:pqts5-rss4}
\end{center}
We can read off the properties of the $5^2_2$-brane when $[p,q] = [0,1]$ in Figure \ref{fig:pqts5-rss4}.
A wrapped NS5-brane is transformed to a bound state of a wrapped NS5-brane and a wrapped KK5-brane.
The Hanany-Witten transition tells us that the wrapped KK5-brane is created from the $5^2_2$-brane.

The third configuration is given by the ST$_{34}$-dualities of Figure \ref{fig:pqs7-rss5}.
In this case, the original $(r,s)_5$-brane is dualized to a bound state of 
$r$ D3-branes and $s$ D5-branes wrapped on the compact two-torus $T^2_{34}$.
In the small size limit of $T^2_{34}$, the new $(r,s)_5$-brane can be seen as a three-brane.
Then we refer to the new one as the $(r,s)_3$-brane.
The configuration is represented in Figure \ref{fig:pqts5-rss3}:
\begin{center}
\slb{.68}{\includegraphics[bb=0 0 161 95]{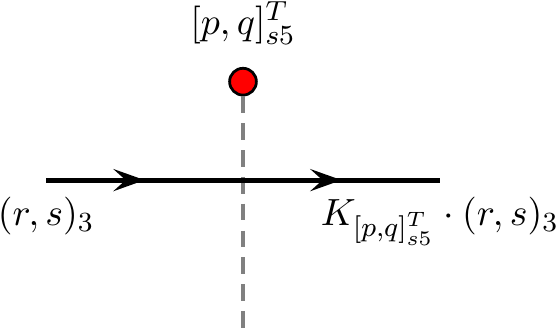}}
\raisebox{12mm}{\slb{.85}{$\xrightarrow{\text{HW}}$}} 
\slb{.68}{\includegraphics[bb=0 0 161 113]{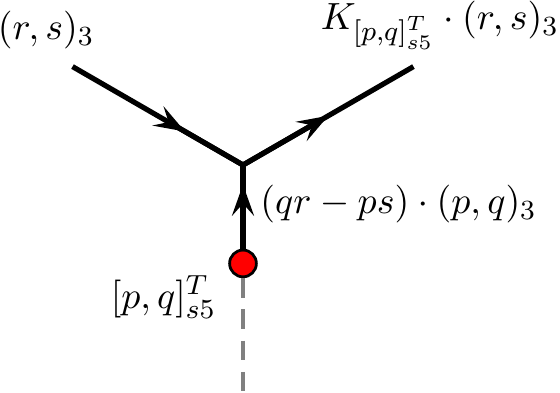}}
\raisebox{13mm}{
\slb{.68}{
\renewcommand{\arraystretch}{1.1}
\begin{tabular}{c:c||c|cc|cc|ccc|c:c} \hline
\multicolumn{2}{c||}{IIB} & 0 & 1 & 2 & \txc{3} & \txc{4} & 5 & 6 & 7 & 8 & 9
\\ \hline\hline
\multirow{2}{*}{$[p,q]^T_{s5}$} 
& $p$ NS5 & $-$ & $-$ & $-$ & & & $-$ & $-$ & $-$ & & 
\\
& $q$ $5^2_2$ & $-$ & $-$ & $-$ & \tcc{2} & \tcc{2} & $-$ & $-$ & $-$ & & 
\\ \hline
& D3 & $-$ & $-$ & $-$ & & & & & & $-$ &
\\
& ``D5'' & $-$ & $-$ & $-$ & $-$ & $-$ & & & & & $-$ 
\\ \hline
\multirow{2}{*}{$(r,s)_3$} 
& $r$ D3 & $-$ & $-$ & $-$ & & & & & & \multicolumn{2}{c}{\multirow{2}{*}{angle}} 
\\
& $s$ ``D5'' & $-$ & $-$ & $-$ & $-$ & $-$ & & & & \multicolumn{2}{c}{} 
\\ \hline
\end{tabular}
}}
\figcaption{A $[p,q]^T_{s5}$-brane and an $(r,s)_3$-brane crossing the branch cut, and its Hanany-Witten transition. This is ST$_{34}$-dual of Figure \ref{fig:pqs7-rss5}.}
\label{fig:pqts5-rss3}
\end{center}
We also find further properties of the exotic $5^2_2$-brane as a $[0,1]^T_{s5}$-brane in Figure \ref{fig:pqts5-rss3}.
This figure emphasizes that a D3-brane is transformed to a bound state of a D3-brane and a wrapped D5-brane by the monodromy.
This wrapped D5-brane is created from the $5^2_2$-brane via the Hanany-Witten transition because of the charge conservation.

We have discussed the new brane junctions associated with the five-brane junction in F-theory. 
Thanks to the duality transformations, we found that the D3-brane wrapped on the compact two-torus ends on the exotic $5^2_2$-brane.
We also found the objects sensitive to the monodromy branch cut of the exotic $5^2_2$-brane when $[p,q] = [0,1]$.
Furthermore, we also understood the brane creations from the exotic $5^2_2$-brane via the Hanany-Witten transitions. 
Utilizing the results which we obtained in this section,
we will study new brane configurations which give rise to the superconformal field theories in the strong gauge coupling limit.

\section{Web of branes with multiple exotic five-branes}
\label{sect:Web-SCFT}

We apply the brane junctions discussed in the previous section to the webs of branes in the presence of multiple exotic branes.
This scenario is associated with the web of five-branes with seven-branes 
in the various works \cite{DeWolfe:1999hj, Kim:2015jba}.
In this section, we find conformal field theories with eight supercharges in five, four, and three dimensions.
For simplicity, we focus only on the brane configurations which yield an $SU(2)$ gauge symmetry on the web of branes,
though it is also possible to consider brane configurations which provide higher-rank gauge symmetries as discussed in \cite{Benini:2009gi, Kim:2012gu, Kim:2015jba}.

Before starting the discussions, 
we introduce the following $[p,q]^T_{s5}$-branes and the corresponding monodromy matrices $K$ analogous to those (\ref{ABC-branes}) in F-theory:
\bsubeq \label{ABC-branes-ts5}
\begin{alignat}{2}
\text{\A-brane} \ &= \ [1,0]^T_{s5} 
\, , &\ls
K_A \ &= \ 
\left(
{\renewcommand{\arraystretch}{.85}
\begin{array}{cc}
1 & -1 
\\
0 & 1
\end{array}
}
\right)
\, , \label{A-brane-ts5} \\
\text{\B-brane} \ &= \ [1,-1]^T_{s5} 
\, , &\ls
K_B \ &= \ 
\left(
{\renewcommand{\arraystretch}{.85}
\begin{array}{cc}
0 & -1 
\\
1 & 2
\end{array}
}
\right)
\, , \label{B-brane-ts5} \\
\text{\C-brane} \ &= \ [1,1]^T_{s5} 
\, , &\ls
K_C \ &= \ 
\left(
{\renewcommand{\arraystretch}{.85}
\begin{array}{cc}
2 & -1 
\\
1 & 0
\end{array}
}
\right)
\, . \label{C-brane-ts5}
\end{alignat}
\esubeq

\subsection{Web of five-branes and 5D SCFTs}
\label{sect:5SCFT}

We begin with the web of $(r,s)_5$-branes with multiple exotic five-branes whose constituents are given in Figure \ref{fig:pqts5-rss5}.
Analogous to the discussion in section \ref{sect:multi7}, we construct a five-brane quadrilateral whose horizontal and vertical segments are $(0,1)_5$-branes (i.e., wrapped $7_3$-branes) and $(1,0)_5$-branes ($5^2_3$-branes), respectively.
Now the distance of two vertical five-branes is measured as $L$.
Because of the charge conservation of the $(r,s)_5$-branes, 
there are four external legs along the $(1,-1)$-direction and the $(1,1)$-direction.
Without changing the physics on the web of five-branes, we can attach the $\B$- and $\C$-branes at the end of these external legs, respectively.
Finally, we put $n$ coincident $\A$-branes (NS5-branes) inside the quadrilateral.
This configuration is expressed in Figure \ref{fig:web5-7-Nf1-ST67}:
\begin{center}
\slb{.8}{\includegraphics[bb=0 0 143 143]{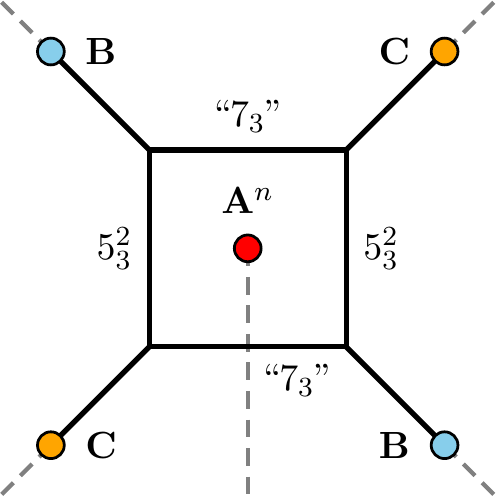}}
\LS
\raisebox{20mm}{
\slb{.8}{
\renewcommand{\arraystretch}{1.1}
\begin{tabular}{c:c||c|cccc|c|cc|c:c} \hline
\multicolumn{2}{c||}{IIB} & 0 & 1 & 2 & 3 & 4 & 5 & \txc{6} & \txc{7} & 8 & 9
\\ \hline\hline
\multirow{2}{*}{$\A^n$} 
& $n$ NS5 & $-$ & $-$ & $-$ & $-$ & $-$ & $-$ & & & & 
\\
& 0 $5^2_2$ & $-$ & $-$ & $-$ & $-$ & $-$ & $-$ & \tcc{2} & \tcc{2} & & 
\\ \hline
& ``$7_3$'' & $-$ & $-$ & $-$ & $-$ & $-$ & & $-$ & $-$ & $-$ &
\\
& $5^2_3$ & $-$ & $-$ & $-$ & $-$ & $-$ & & \tcc{2} & \tcc{2} & & $-$
\\ \hline
\multirow{2}{*}{$(r,s)_5$} 
& $r$ ``$7_3$'' & $-$ & $-$ & $-$ & $-$ & $-$ & & $-$ & $-$ & \multicolumn{2}{c}{\multirow{2}{*}{angle}} 
\\
& $s$ $5^2_3$ & $-$ & $-$ & $-$ & $-$ & $-$ & & \tcc{2} & \tcc{2} & \multicolumn{2}{c}{} 
\\ \hline
\end{tabular}
}
}
\figcaption{A five-brane quadrilateral with $n$ coincident $\A$-branes, two $\B$-branes and two $\C$-branes. 
The table expresses the extending directions of branes.
This brane configuration induces the 5D $\N=1$ $SU(2)$ gauge theory with $n$ flavors along the 01234-directions on the horizontal wrapped $7_3$-branes.}
\label{fig:web5-7-Nf1-ST67}
\end{center}
This configuration preserves 1/4-BPS and all the branes have the common 01234-directions in their worldvolumes.
Moving the $\A^n$-branes horizontally, we can find that $n$ horizontal $7_3$-branes which end on the vertical $5^2_3$-branes are created by the Hanany-Witten transitions.
Hence, in the gravity decoupling limit, the five-dimensional $\N=1$ $SU(2)$ gauge theory with $n$ flavors emerges on the 01234-directions of the $7_3$-branes.
However, we should notice that the gauge fields and the matters originate from the excitation modes of open D-strings ending on the horizontal $7_3$-branes rather than F-strings.
The W-boson is given by an open D-string stretched between two horizontal $7_3$-branes. 
The instanton comes from the open D3-branes wrapped on the two-torus $T^2_{67}$ and stretched between two vertical $5^2_3$-branes (see section \ref{sect:ts5-mono}).
It is remarked that we can also derive this configuration from Figure \ref{fig:web5-7} via the ST$_{67}$-dualities, where the size of the compact two-torus $T^2_{67}$ should be infinitesimal.

Following the discussion in \cite{DeWolfe:1999hj}, we consider the strong gauge coupling limit of the gauge theory on the five-branes.
To do so, we first evaluate the gauge coupling constant $g_{\text{YM}}$ on the horizontal $7_3$-branes.
Although we do not know the DBI action of the $7_3$-brane rigorously,
we can estimate it from that of the D7-brane via the S-duality.
The DBI action of the single D$p$-brane is 
\begin{align}
S_{\text{D$p$}} \ &= \ 
\frac{1}{g_s \l_s^{p+1}} \int \d^{p+1} \xi \, \e^{-\phi} \, \sqrt{- \det (g_{ab} + B_{ab} + \l_s^2 F_{ab})}
\, , \label{DBI-Dp}
\end{align}
where $\l_s$ is the string length and $\xi$ is the worldvolume coordinates.
$\phi$, $g_{ab}$ and $B_{ab}$ are the dilaton, the induced metric and the pull-back of the NS-NS B-field, respectively.
$F_{ab}$ is the field strength of the worldvolume gauge potential.
Since we are interested in the gauge coupling constant $g_{\text{YM}}$ from the D7 action, we extract the quadratic term of $F_{ab}$ around the flat background:
\begin{align}
S_{\text{D7}}
\ &= \ 
\frac{1}{g_s \l_s^8} \int \d^8 \xi \, \Big\{
\l_s^4 F_{ab}^2 
+ \ldots
\Big\}
\, . 
\end{align}
Here we generally ignored numerical factors.
Now we estimate the DBI action of the wrapped $7_3$-brane in Figure \ref{fig:web5-7-Nf1-ST67}.
We perform the S-duality by replacing $g_s \to 1/g_s$ and $\l_s^2 \to g_s \l_s^2$, and the worldvolume integral $\int \d^8 \xi$ is given as
\begin{align}
(2 \pi \wt{R}_6) (2 \pi \wt{R}_7) L \int \d^5 \xi
\, ,
\end{align}
where $\wt{R}_i$ are the radii of the compact two-torus $T^2_{67}$ after the T-dualities. 
Then we obtain the relation between the gauge coupling constant $g_{\text{YM}}$ and the parameters on the wrapped $7_3$-branes:
\begin{align}
\frac{1}{g_{\text{YM}}^2}
\ &= \ 
\frac{L}{g_s}
\Big( \frac{2 \pi \wt{R}_6}{\l_s^2} \Big)
\Big( \frac{2 \pi \wt{R}_7}{\l_s^2} \Big)
\, . \label{gYM-73}
\end{align}
In the gravity decoupling and the strong gauge coupling limit,
we should take the following limit:
\begin{align}
\l_s \to 0
\, , \ls
\frac{\wt{R}_i}{\l_s^2} \to 0
\, , \ls
L \to 0
\, . \label{strongYM-w73}
\end{align}
This is the small size limit of the quadrilateral in Figure \ref{fig:web5-7-Nf1-ST67}. 
We also emphasize that the radii of the compact circles $R_i = \l_s^2/\wt{R}_i$ before the T-dualities, i.e., the size of radii in Figure \ref{fig:web5-7}, are infinitely large.
Since the gauge theory lives in the 01234-directions, 
the strongly coupled theory emerges at the UV fixed point.
Hence, in the same way as in \cite{DeWolfe:1999hj},
the superconformal field theory with $n$ flavors under the global $E_{n+1}$ symmetry should be non-trivially realized.
We qualitatively argue this in terms of the five-brane web illustrated in Figure \ref{fig:F-HW-small-En-ST67}:
\begin{center}
\slb{.8}{\includegraphics[bb=0 0 143 143]{./Anbox-ST67.pdf}}
\raisebox{18mm}{$\xrightarrow{\text{HW}}$} 
\raisebox{10mm}{\slb{.8}{\includegraphics[bb=0 0 97 69]{./AnBCBC.pdf}}}
\raisebox{18mm}{$\xrightarrow{\text{reordering}}$}  
\raisebox{10mm}{\slb{.8}{\includegraphics[bb=0 0 97 69]{./AnBCCX.pdf}}}
\raisebox{18mm}{$\xrightarrow{\text{small}}$} 
\raisebox{16mm}{\slb{.8}{\includegraphics[bb=0 0 77 37]{./EnX.pdf}}}
\figcaption{After the Hanany-Witten transitions of Figure \ref{fig:web5-7-Nf1-ST67}, we reorder the five-branes by moving their branch cuts. Further we take the small size limit of the five-brane loop when $n = 5,6,7$.}
\label{fig:F-HW-small-En-ST67}
\end{center}
We begin with the five-brane quadrilateral in the presence of $n$ coincident $\A$-branes, two $\B$-branes and two $\C$-branes.
Through the Hanany-Witten transitions, all of the branes are put inside the quadrilateral.
Then the four external legs are annihilated and the quadrilateral is highly curved by the back reactions from the $\A^n\B\C\B\C$-branes.
We refer to the deformed quadrilateral as the five-brane loop.
By moving the monodromy branch cuts inside the loop, 
the ordering of the branes becomes $\A^n \B \C \C \X_{[3,1]}$, 
where $\X_{[3,1]}$ is the exotic $[3,1]^T_{s5}$-brane.
We find that the $\A^n \B \C \C$-branes are collapsible and becomes the single $\E_{n+1}$-brane as far as we concern $n = 5,6,7$.
In the strong gauge coupling limit (\ref{strongYM-w73}), 
the five-brane loop shrinks to zero size.
Since the $\E_{n+1}$-brane and the $\X_{[3,1]}$-brane are not collapsible, 
the $\X_{[3,1]}$-brane goes outside and moves infinitely far away from the loop, 
though the $\E_{n+1}$-brane remains inside the loop.
Then, in the limit, 
the five-dimensional SCFT appears.
The open D-strings stretched between the $\E_{n+1}$-brane and the loop provide the global $E_{n+1}$ symmetry.
The quantitative analysis is completely parallel to that of the ordinary five-brane web \cite{DeWolfe:1999hj}.
This discussion is also applicable even if $0 \leq n \leq 4$ and $n = 8$ \cite{Kim:2015jba}.
We conclude that the five-dimensional $\N=1$ SCFTs with $E_{n+1}$ symmetry can be realized in the $[p,q]^T_{s5}$-branes background.

\subsection{Web of four-branes and 4D SCFTs}
\label{sect:4SCFT}

Next, we consider the web of $(r,s)_4$-branes in the presence of the exotic five-branes (\ref{ABC-branes-ts5}), whose building blocks are discussed in Figure \ref{fig:pqts5-rss4}.
We set the four-brane quadrilateral with $\A^n$-, $\B$- and $\C$-branes as in Figure \ref{fig:web5-7-Nf1-ST47}.
\begin{figure}[h]
\begin{center}
\slb{.8}{\includegraphics[bb=0 0 143 143]{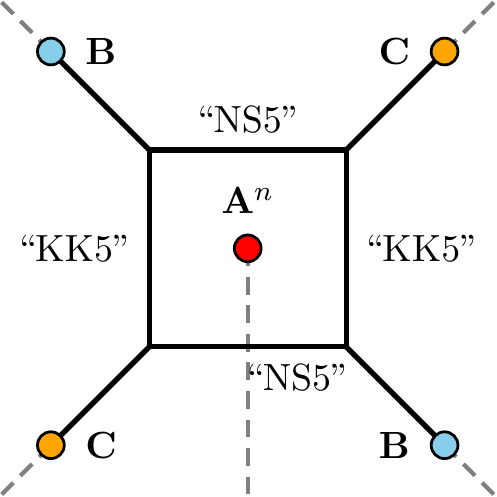}}
\LS
\raisebox{20mm}{
\slb{.8}{
\renewcommand{\arraystretch}{1.1}
\begin{tabular}{c:c||c|ccc|c|cc|c|c:c} \hline
\multicolumn{2}{c||}{IIB} & 0 & 1 & 2 & 3 & \txc{4} & 5 & 6 & \txc{7} & 8 & 9
\\ \hline\hline
\multirow{2}{*}{$\A^n$} 
& $n$ NS5 & $-$ & $-$ & $-$ & $-$ & & $-$ & $-$ & & & 
\\
& 0 $5^2_2$ & $-$ & $-$ & $-$ & $-$ & \tcc{2} & $-$ & $-$ & \tcc{2} & & 
\\ \hline
& ``NS5'' & $-$ & $-$ & $-$ & $-$ & & & & $-$ & $-$ &
\\
& ``KK5'' & $-$ & $-$ & $-$ & $-$ & $-$ & & & \tcc{2} & & $-$
\\ \hline
\multirow{2}{*}{$(r,s)_4$} 
& $r$ ``NS5'' & $-$ & $-$ & $-$ & $-$ & & & & $-$ & \multicolumn{2}{c}{\multirow{2}{*}{angle}} 
\\
& $s$ ``KK5'' & $-$ & $-$ & $-$ & $-$ & $-$ & & & \tcc{2} & \multicolumn{2}{c}{} 
\\ \hline
\end{tabular}
}
}
\figcaption{A four-brane quadrilateral with $n$ coincident $\A$-branes, two $\B$-branes and two $\C$-branes. This brane configuration induces the 4D $\N=2$ $SU(2)$ gauge theory with $n$ flavors along the 0123-directions on the horizontal NS5-branes. This is ST$_{47}$-dual of Figure \ref{fig:web5-7}.}
\label{fig:web5-7-Nf1-ST47}
\end{center}
\end{figure}
In this configuration,
the gauge theory with eight supercharges emerges on two horizontal wrapped NS5-branes.
The W-boson and the instanton come from an open D-string stretched between two horizontal wrapped NS5-branes and an open D3-brane wrapped on the two-torus $T^2_{47}$ ending on two vertical wrapped KK5-branes, respectively.
In a similar fashion,
we investigate the strong gauge coupling limit of the theories with eight supercharges on the four-brane web.
First, we estimate the gauge coupling constant on the worldvolume of the horizontal wrapped NS5-branes.
The DBI action of the single NS5-brane is obtained from that of the D$p$-brane (\ref{DBI-Dp}) via the string dualities\footnote{Of course the DBI action of the single NS5-brane is well known. For instance, see \cite{Kimura:2014upa} in which the DBI actions of the NS5-brane, the D5-brane, the exotic $5^2_2$-brane and the exotic $5^2_3$-brane are exhibited.}.
Performing the ST$_{67}$-dualities, 
we can extract the kinetic term of the gauge potential on the wrapped NS5-brane:
\bsubeq \label{gYM-wNS5}
\begin{gather}
S_{\text{NS5}} 
\ = \ 
\frac{1}{g_s^2 \l_s^6} \int \d^6 \xi \, \Big\{
(g_s \l_s^2)^2 F_{ab}^2
+ \ldots
\Big\}
\ = \ 
\frac{(2 \pi \wt{R}_7) L}{\l_s^2} \int \d^4 \xi \, F_{ab}^2 
+ \ldots
\, , \\
\therefore \ \ \ 
\frac{1}{g_{\text{YM}}^2}
\ = \ 
\frac{(2 \pi \wt{R}_7) L}{\l_s^2} 
\, . 
\end{gather}
\esubeq
Here $\wt{R}_7$ is the radius of the compact 7th-direction after the T-duality.
Then the strong gauge coupling limit is given as
\begin{align}
\l_s \to 0
\, , \ls
\frac{\wt{R}_7}{\l_s^2} \to 0
\, , \ls
L \to 0
\, . \label{strongYM-wNS5}
\end{align}
For example, we qualitatively discuss the gauge theory and its strong coupling limit in the case of $n = 5,6,7$ in Figure \ref{fig:F-HW-small-En-ST47}:
\begin{center}
\slb{.8}{\includegraphics[bb=0 0 143 143]{./Anbox-ST47.pdf}}
\raisebox{18mm}{$\xrightarrow{\text{HW}}$} 
\raisebox{10mm}{\slb{.8}{\includegraphics[bb=0 0 97 69]{./AnBCBC.pdf}}}
\raisebox{18mm}{$\xrightarrow{\text{reordering}}$} 
\raisebox{10mm}{\slb{.8}{\includegraphics[bb=0 0 97 69]{./AnBCCX.pdf}}}
\raisebox{18mm}{$\xrightarrow{\text{small}}$} 
\raisebox{16mm}{\slb{.8}{\includegraphics[bb=0 0 77 37]{./EnX.pdf}}}
\figcaption{After the Hanany-Witten transitions of Figure \ref{fig:web5-7-Nf1-ST47}, we reorder the five-branes by moving their branch cuts. Further we take the small size limit of the four-brane loop when $n = 5,6,7$.}
\label{fig:F-HW-small-En-ST47}
\end{center}
Following the discussions of Figures \ref{fig:F-HW-small-En} and \ref{fig:F-HW-small-En-ST67}, we find the global $E_{n+1}$ symmetry in the strong coupling limit.
This is realized at the non-trivial fixed point discussed in \cite{Minahan:1996fg, Minahan:1996cj}.
Hence we found that the four-dimensional $\N=2$ SCFTs with $n$ flavors ($0 \leq n \leq 7$) are also obtained by the exotic $[p,q]^T_{s5}$-branes, though
the field theory contents originated from the excitations of the four-branes are different from the original five-brane web. 
We can also think of the Tao diagram \cite{Kim:2015jba} when $n = 8$ in this four-brane web in the presence of the exotic five-branes.

\subsection{Web of three-branes and 3D SCFTs}
\label{sect:3SCFT}

Finally, we investigate the web of $(r,s)_3$-branes in the presence of the multiple exotic $[p,q]^T_{s5}$-branes whose constituents are given by Figure \ref{fig:pqts5-rss3}.
Again, we consider a simple model given by a quadrilateral with exotic branes illustrated in Figure \ref{fig:web5-7-Nf1-ST34}:
\begin{center}
\slb{.8}{\includegraphics[bb=0 0 143 143]{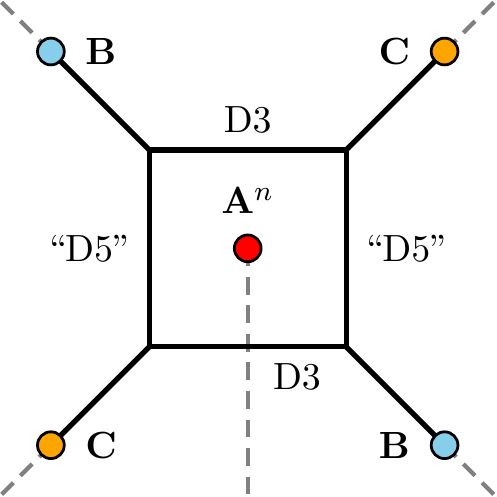}}
\LS
\raisebox{20mm}{
\slb{.8}{
\renewcommand{\arraystretch}{1.1}
\begin{tabular}{c:c||c|cc|cc|ccc|c:c} \hline
\multicolumn{2}{c||}{IIB} & 0 & 1 & 2 & \txc{3} & \txc{4} & 5 & 6 & 7 & 8 & 9
\\ \hline\hline
\multirow{2}{*}{$\A^n$} 
& 1 NS5 & $-$ & $-$ & $-$ & & & $-$ & $-$ & $-$ & & 
\\
& 0 $5^2_2$ & $-$ & $-$ & $-$ & \tcc{2} & \tcc{2} & $-$ & $-$ & $-$ & & 
\\ \hline
& D3 & $-$ & $-$ & $-$ & & & & & & $-$ &
\\
& ``D5'' & $-$ & $-$ & $-$ & $-$ & $-$ & & & & & $-$
\\ \hline
\multirow{2}{*}{$(r,s)_3$} 
& $r$ D3 & $-$ & $-$ & $-$ & & & & & & \multicolumn{2}{c}{\multirow{2}{*}{angle}} 
\\
& $s$ ``D5'' & $-$ & $-$ & $-$ & $-$ & $-$ & & & & \multicolumn{2}{c}{} 
\\ \hline
\end{tabular}
}
}
\figcaption{A three-brane quadrilateral with $n$ coincident $\A$-branes, two $\B$-branes and two $\C$-branes. This brane configuration induces the 3D $\N=4$ $SU(2)$ gauge theory with $n$ flavors along the 012-directions on the horizontal D3-branes. This configuration is derived from Figure \ref{fig:web5-7} via the string ST$_{34}$-dualities.}
\label{fig:web5-7-Nf1-ST34}
\end{center}
In this brane configuration, 
the three-dimensional $\N=4$ gauge theory with $n$ flavors can be realized along the 012-directions.
The W-boson and the instanton in the gauge theory originate from open D-strings stretched with two horizontal D3-branes and open D3-branes wrapped on $T^2_{34}$ ending on two vertical wrapped D5-branes, respectively.

The gauge coupling on the horizontal D3-branes can be obtained by the DBI action (\ref{DBI-Dp}) in such a way that
\bsubeq \label{gYM-D3}
\begin{gather}
S_{\text{D3}} 
\ = \ 
\frac{1}{g_s \l_s^4} \int \d^4 \xi \, \Big\{
\l_s^4 F_{ab}^2
+ \ldots
\Big\}
\ = \ 
\frac{L}{g_s} \int \d^3 \xi \, F_{ab}^2 
+ \ldots
\, , \\
\therefore \ \ \ 
\frac{1}{g_{\text{YM}}^2}
\ = \ 
\frac{L}{g_s} 
\, . 
\end{gather}
\esubeq
Then the strong gauge coupling limit is nothing but the small size limit:
\begin{align}
L \ \to \ 0
\, . \label{strongYM-D3}
\end{align}
Analogous to the previous discussions, we can again qualitatively discuss the emergence of the superconformal theory illustrated in Figure \ref{fig:F-HW-small-En-ST34}:
\begin{center}
\slb{.8}{\includegraphics[bb=0 0 143 143]{./Anbox-ST34.pdf}}
\raisebox{18mm}{$\xrightarrow{\text{HW}}$} 
\raisebox{10mm}{\slb{.8}{\includegraphics[bb=0 0 97 69]{./AnBCBC.pdf}}}
\raisebox{18mm}{$\xrightarrow{\text{reordering}}$} 
\raisebox{10mm}{\slb{.8}{\includegraphics[bb=0 0 97 69]{./AnBCCX.pdf}}}
\raisebox{18mm}{$\xrightarrow{\text{small}}$} 
\raisebox{16mm}{\slb{.8}{\includegraphics[bb=0 0 77 37]{./EnX.pdf}}}
\figcaption{After the Hanany-Witten transitions of Figure \ref{fig:web5-7-Nf1-ST34}, we reorder the five-branes by moving their branch cuts. Further we take the small size limit of the three-brane loop when $n = 5,6,7$.}
\label{fig:F-HW-small-En-ST34}
\end{center}
Even though the dimensions of the brane web is different from the others,
we obtain the collapsible $\E_{n+1}$-brane inside the loop and the $\X_{[3,1]}$-brane going infinitely far away from the loop in the strong coupling limit.
Hence, we find that three-dimensional $\N=4$ $SU(2)$ gauge theories with $n = 5,6,7$ flavors flow to the $\N=4$ SCFTs with $E_{n+1}$ symmetry.

\section{Summary and discussions}
\label{sect:summary}

In this paper, we studied various brane junctions and their Hanany-Witten transitions originated from the exotic branes.
Applying the string dualities to the $[p,q]$ 7-brane in F-theory, we found the monodromy matrices of various $[p,q]$-branes in lower dimensions.
In particular, we carefully investigated various configurations involving the exotic $5^2_2$-brane.
Furthermore, we analyzed the web of branes in the presence of the multiple exotic five-branes analogous to the web of five-branes with multiple seven-branes in F-theory.
We obtained an exotic brane configuration which provides the same SCFTs in five dimensions.
We also obtained new configurations with the exotic five-branes which give rise to non-trivial SCFTs with enhanced $E_{n+1}$ symmetry in four and three dimensions. 

There are open questions.
We can formally construct the SCFTs with $n = 8$ flavors when we put eight coincident $\A$-branes inside the web of branes.
In five-dimensional case, the SCFT describes the E-string as discussed in \cite{Kim:2015jba}.
In four and three dimensions, we can also construct the brane configurations including eight coincident $\A$-branes.
It will be interesting if their SCFTs describe new stringy phenomena.
We did not explicitly described other exotic $[p,q]$-branes, although they are similarly found as in the case of the exotic $[p,q]^T_{s5}$-branes.
In appendix \ref{app:ex-EBJ}, we briefly introduce one example.


We have studied the property of the exotic branes in the conventional frameworks.
It is also known that $\beta$-supergravity and its U-duality extension \cite{Andriot:2013xca, Andriot:2014uda, Blair:2014zba, Sakatani:2014hba} will provide much richer configurations of various branes in globally
nongeometric backgrounds\footnote{We note that the U-duality extended version of $\beta$-supergravity \cite{Blair:2014zba, Sakatani:2014hba} involves the Ramond-Ramond potentials $C_p$ and their string dualized objects $\gamma^p$. This formulation may be referred to as ``$\gamma$-supergravity''.}.
For instance, there exist further $SL(2,{\mathbb Z})_S$ S-duality doublets in Figure \ref{fig:E-dualities}.
A pair of the $5^2_3$-brane and the $5^2_2$-brane is a doublet analogous to the ordinary pair of the D5-brane and the NS5-brane.
A pair of the $1^6_4$-brane and the $1^6_3$-brane is also a doublet.
This is similar to a pair of the F-string and the D-string.
However, such pairs of exotic branes never appear in the conventional frameworks. 
They would be doublets in the nongeometric backgrounds and could be captured by the above extended versions of the ordinary supergravity.

Here we exhibit two configurations of the brane junctions in the nongeometric backgrounds.
One is the five-branes junction involving the pair of the $5^2_3$-brane and the $5^2_2$-brane rather than the pair of the D5-brane and the NS5-brane. See Table \ref{fig:pqs7-rsse5-8DB}:
\begin{center}
\slb{.8}{
\renewcommand{\arraystretch}{1.1}
\begin{tabular}{c:c||c|cccc|c|cc|c:c} \hline
\multicolumn{2}{c||}{IIB} & 0 & 1 & 2 & 3 & 4 & 5 & \txc{6} & \txc{7} & 8 & 9
\\ \hline\hline
\multirow{2}{*}{$[p,q]^S_{7}$} 
& $p$ ``D7'' & $-$ & $-$ & $-$ & $-$ & $-$ & $-$ & $-$ & $-$ & & 
\\
& $q$ ``$7_3$'' & $-$ & $-$ & $-$ & $-$ & $-$ & $-$ & $-$ & $-$ & & 
\\ \hline
& $5^2_3$ & $-$ & $-$ & $-$ & $-$ & $-$ & & \tcc{2} & \tcc{2} & $-$ &
\\
& $5^2_2$ & $-$ & $-$ & $-$ & $-$ & $-$ & & \tcc{2} & \tcc{2} & & $-$ 
\\ \hline
\multirow{2}{*}{$(r,s)_5$} 
& $r$ $5^2_3$ & $-$ & $-$ & $-$ & $-$ & $-$ & & \tcc{2} & \tcc{2} & \multicolumn{2}{c}{\multirow{2}{*}{angle}} 
\\
& $s$ $5^2_2$ & $-$ & $-$ & $-$ & $-$ & $-$ & & \tcc{2} & \tcc{2} & \multicolumn{2}{c}{} 
\\ \hline
\end{tabular}
}
\tbcaption{A $(r,s)_5$-brane, as a vortex, crossing the branch cut of $[p,q]^S_{7}$-brane in 8D, where the 67-directions are compactified.}
\label{fig:pqs7-rsse5-8DB}
\end{center}
This configuration is never realized in ten dimensions because the 67-directions are compactified.
This cannot be obtained from any brane configurations in the conventional supergravity theories via the string dualities.
Furthermore, in the current understanding, it is unclear what kind of objects are ending on the $[p,q]^S_7$-brane and this nongeometric $(r,s)_5$-brane simultaneously.
A similar situation occurs when we consider the second configuration expressed in Table \ref{fig:pqs7-rsse1-4DB}:
\begin{center}
\slb{.8}{
\renewcommand{\arraystretch}{1.1}
\begin{tabular}{c:c||c|c|cccccc|c:c} \hline
\multicolumn{2}{c||}{IIB} & 0 & 1 & \txc{2} & \txc{3} & \txc{4} & \txc{5} & \txc{6} & \txc{7} & 8 & 9
\\ \hline\hline
\multirow{2}{*}{$[p,q]^S_{7}$} 
& $p$ ``D7'' & $-$ & $-$ & $-$ & $-$ & $-$ & $-$ & $-$ & $-$ & & 
\\
& $q$ ``$7_3$'' & $-$ & $-$ & $-$ & $-$ & $-$ & $-$ & $-$ & $-$ & & 
\\ \hline
& $1^6_4$ & $-$ & & \tcc{2} & \tcc{2} & \tcc{2} & \tcc{2} & \tcc{2} & \tcc{2} & $-$ &
\\
& $1^6_3$ & $-$ & & \tcc{2} & \tcc{2} & \tcc{2} & \tcc{2} & \tcc{2} & \tcc{2} & & $-$ 
\\ \hline
\multirow{2}{*}{$(r,s)_1$} 
& $r$ $1^6_4$ & $-$ & & \tcc{2} & \tcc{2} & \tcc{2} & \tcc{2} & \tcc{2} & \tcc{2} & \multicolumn{2}{c}{\multirow{2}{*}{angle}} 
\\
& $s$ $1^6_3$ & $-$ & & \tcc{2} & \tcc{2} & \tcc{2} & \tcc{2} & \tcc{2} & \tcc{2} & \multicolumn{2}{c}{} 
\\ \hline
\end{tabular}
}
\tbcaption{A $(r,s)$-string, as a vortex, crossing the branch cut of $[p,q]^S_{7}$-brane in 4D, where the 234567-directions are compactified.}
\label{fig:pqs7-rsse1-4DB}
\end{center}
This resembles the configuration in Figures \ref{fig:pqs7-rss1} and \ref{fig:pqs7-rss1-HW}. 
However, this is realized in four-dimensional spacetime.
This configuration is not related to any brane configurations in the framework of the conventional supergravity, neither.
The above two configurations are the typical ones which should be understood in $\beta$- or $\gamma$-supergravity.
It would also be important that the exotic $[p,q]$-branes should be represented in the framework of exceptional field theories (or EFTs, for short).
EFTs in diverse dimensions have been already constructed as in Table \ref{table:EFTs}:
\begin{center}
\slb{.85}{$\renewcommand{\arraystretch}{1.3}
\begin{array}{c:c|rcl|c||c:c|c|c} \hline
D & p & \multicolumn{3}{c|}{\text{U-duality group}} & \text{EFT}
& 
D & p & \text{U-duality group}  & \text{EFT}
\\ \hline\hline
9 & 6 & E_{2(2)}({\mathbb Z}) \!\!&=&\!\! SL(2,{\mathbb Z}) \times {\mathbb Z}_2 & \cite{Berman:2015rcc} &
5 & 2 & E_{6(6)}({\mathbb Z}) & \cite{Hohm:2013vpa, Musaev:2014lna}
\\
8 & 5 & E_{3(3)}({\mathbb Z}) \!\!&=&\!\! SL(3,{\mathbb Z}) \times SL(2,{\mathbb Z}) & \cite{Hohm:2015xna} &
4 & 1 & E_{7(7)}({\mathbb Z}) & \cite{Hohm:2013uia}
\\
7 & 4 & E_{4(4)}({\mathbb Z}) \!\!&=&\!\! SL(5,{\mathbb Z}) & \cite{Park:2013gaj, Blair:2014zba, Musaev:2015ces} &
3 & 0 & E_{8(8)}({\mathbb Z}) & \cite{Hohm:2014fxa}
\\ \cdashline{7-10}
6 & 3 & E_{5(5)}({\mathbb Z}) \!\!&=&\!\! SO(5,5;{\mathbb Z}) & \cite{Abzalov:2015ega} &
11-n & 8-n & SL(n+1;{\mathbb Z}) \subseteq E_{n(n)}({\mathbb Z}) & \cite{Park:2014una} 
\\ \hline
\end{array}
$}
\tbcaption{U-duality groups and the corresponding EFTs. $D$ and $p$ denote the spacetime dimensions and the spatial dimensions of defect branes, respectively.}
\label{table:EFTs}
\end{center}
Once the exotic branes are analyzed in EFTs, we will be able to understand all of the exotic structures of string theory beyond the $SL(2,{\mathbb Z})$ duality subgroup of the full U-duality symmetry groups, as proposed in \cite{Kumar:1996zx, Lu:1998sx}.

\vspace{3mm}

We conclude that, as far as we concern the configurations derived from the $[p,q]$ 7-branes in F-theory, 
the techniques we utilized in the configurations with the exotic $5^2_2$-branes can also be applicable to {\it any} configurations in the presence of defect branes in {\it any} spacetime dimensions. 
Hence, applying the techniques of the seven-branes to the exotic branes, 
we can construct F-theories in diverse dimensions.
When we understand the whole structures of EFTs, we can construct further generalized F-theories involving the highly nongeometric configurations in Tables \ref{fig:pqs7-rsse5-8DB} and \ref{fig:pqs7-rsse1-4DB} in various dimensions.

\section*{Acknowledgments}

I would like to thank 
Yosuke Imamura,
Hirotaka Kato, 
Hiroki Matsuno, 
Takahiro Nishinaka,
Masaki Shigemori,
Hongfei Shu,
Shigeki Sugimoto
and 
Masato Taki
for helpful discussions.
I also thank the Yukawa Institute for Theoretical Physics at Kyoto University for hospitality during the YITP workshop on ``Microstructures of black holes'' ({YITP-W-15-20}).
I am supported by the MEXT-Supported Program for the Strategic Research Foundation at Private Universities ``Topological Science'' ({Grant No.~S1511006}). 
I am also supported in part by the Iwanami-Fujukai Foundation.

\begin{appendix}
\section*{Appendix}

\section{An example: exotic brane junction by $[p,q]^E_{d3}$-brane}
\label{app:ex-EBJ}

In this appendix, we exhibit an example of the exotic brane junctions involving the single $[p,q]^E_{d3}$-brane.
The $[p,q]^E_{d3}$-brane is an $SL(2,{\mathbb Z})_E$ transformed object of the single D3-brane smeared four of the six transverse directions.
This is also obtained from the single $[p,q]^S_7$-brane via the T-dualities along the four worldvolume directions.
The $(p,q)$-string ending on the $[p,q]^S_7$-brane, the bound state of $p$ F-strings and $q$ D-strings is also transformed to a bound state of $p$ F-strings and $q$ D5-branes wrapped on the compact four-torus $T^4$.
When the size of $T^4$ is infinitesimal, we can regard the new object as a ``string''. Then we continue to refer to it as the $(p,q)$-string.

We consider a brane junction originated from the single $[p,q]^E_{d3}$-brane. 
For instance, we apply the T$_{2347}$-dualities to the configuration in Figure \ref{fig:pqs7-rss5}. We illustrate the resultant configuration in Figure \ref{fig:pqed3-rss2}:
\begin{center}
\slb{.67}{\includegraphics[bb=0 0 161 95]{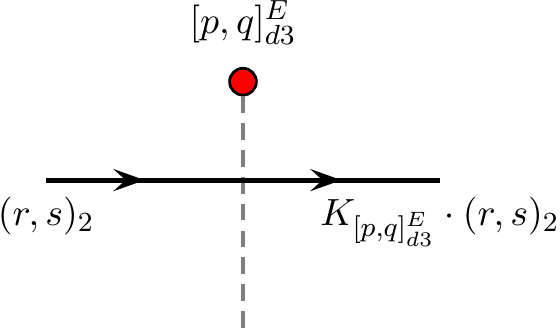}}
\raisebox{12mm}{\slb{.85}{$\xrightarrow{\text{HW}}$}} 
\slb{.67}{\includegraphics[bb=0 0 161 113]{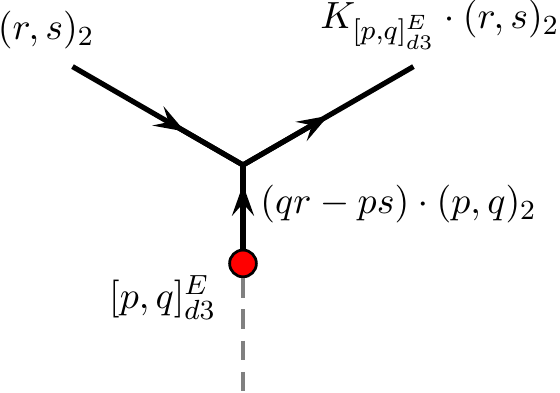}}
\raisebox{13mm}{
\slb{.67}{
\renewcommand{\arraystretch}{1.1}
\begin{tabular}{c:c||c|c|ccc|cc|c|c:c} \hline
\multicolumn{2}{c||}{IIB} & 0 & 1 & \txc{2} & \txc{3} & \txc{4} & 5 & 6 & \txc{7} & 8 & 9
\\ \hline\hline
\multirow{2}{*}{$[p,q]^E_{d3}$} 
& $p$ D3 & $-$ & $-$ & & & & $-$ & $-$ & & & 
\\
& $q$ $3^4_3$ & $-$ & $-$ & \tcc{2} & \tcc{2} & \tcc{2} & $-$ & $-$ & \tcc{2} & & 
\\ \hline
& ``D3'' & $-$ & $-$ & & & & & & $-$ & $-$ &
\\
& ``KK5'' & $-$ & $-$ & $-$ & $-$ & $-$ & & & \tcc{2} & & $-$ 
\\ \hline
\multirow{2}{*}{$(r,s)_2$} 
& $r$ ``D3'' & $-$ & $-$ & & & & & & $-$ & \multicolumn{2}{c}{\multirow{2}{*}{angle}} 
\\
& $s$ ``KK5'' & $-$ & $-$ & $-$ & $-$ & $-$ & & & \tcc{2} & \multicolumn{2}{c}{} 
\\ \hline
\end{tabular}
}
}
\figcaption{A $[p,q]^E_{d3}$-brane and an $(r,s)_2$-brane crossing the branch cut, and its Hanany-Witten transition. This is T$_{2347}$-dual of Figure \ref{fig:pqs7-rss5}. This configuration is self-dual under the S-duality, though the constituents of the $(p,q)$-string ending on the $[p,q]^E_{d3}$-brane is transformed.}
\label{fig:pqed3-rss2}
\end{center}
Because of the T$_{2347}$-dualities, the original $(r,s)_5$-brane in Figure \ref{fig:pqs7-rss5} is transformed to a $(r,s)_2$-brane whose constituents are $r$ D3-branes wrapped on the compact $S^1_7$ along the 7th-direction and $s$ KK5-branes wrapped on the compact three-torus $T^3_{234}$ along the 234-directions.
This configuration is interesting. 
This is because all of the branes, except for the $(p,q)$-string ending on the $[p,q]^E_{d3}$-brane, are self-dual under the string S-duality.
Analogous to the web of five-branes in Figure \ref{fig:web5-7},
if we construct webs of two-branes in which the configuration in Figure \ref{fig:pqed3-rss2} are the building blocks, 
we may obtain two-dimensional SCFTs with non-trivial flavor symmetry and eight supercharges. 
These SCFTs are invariant under a certain duality symmetry associated with the string S-duality.

\end{appendix}

}
\end{document}